\documentclass[twocolumn,showpacs,twoside,10pt,aps,prl,superscriptaddress]{revtex4-2}
\usepackage{amsmath}
\usepackage{makecell}
\usepackage[switch]{lineno}
\usepackage{graphicx}
\usepackage{dcolumn}
\usepackage{bm}
\usepackage[colorlinks=true,linkcolor=blue,citecolor=blue,urlcolor=blue]{hyperref}
\usepackage{orcidlink}
\begin{document}
\preprint{APS/123-QED}
\title{Entanglement-enhanced quantum metrology via alternating in-phase and quadrature modulation}
\def\SZU{Institute of Quantum Precision Measurement, State Key Laboratory of Radio Frequency Heterogeneous Integration, \\
College of Physics and Optoelectronic Engineering, Shenzhen University, Shenzhen 518060, China}
\def\SYSUZH{Laboratory of Quantum Engineering and Quantum Metrology, School of Physics and Astronomy,\\ 
Sun Yat-Sen University (Zhuhai Campus), Zhuhai 519082, China}
\def\GD{Quantum Science Center of Guangdong-Hong Kong-Macao Greater Bay Area (Guangdong), Shenzhen 518045, China}

\author{Jihao Ma~\orcidlink{0009-0003-4068-4590}}
  \affiliation{\SZU}
  \affiliation{\SYSUZH}
  
\author{Jiahao Huang~\orcidlink{0000-0001-7288-9724}}
  \altaffiliation{Email: hjiahao@mail2.sysu.edu.cn, eqjiahao@gmail.com}
  \affiliation{\SZU}
  \affiliation{\SYSUZH}

\author{Chaohong Lee~\orcidlink{0000-0001-9883-5900}}
  \altaffiliation{Email: chleecn@szu.edu.cn, chleecn@gmail.com}
  \affiliation{\SZU}
  \affiliation{\GD}
  
\date{\today}
\begin{abstract}
Quantum metrology harnesses quantum entanglement to improve measurement precision beyond standard quantum limit. 
Although nonlinear interaction is essential for generating entanglement, during signal accumulation, it becomes detrimental and therefore must be suppressed. 
To address this challenge, we propose an alternating in-phase and quadrature modulation (AIQM) scheme, designed to operate under a fixed nonlinear interaction.
During signal accumulation, our time-interleaved approach sequentially applies the in-phase and quadrature driving fields, thereby eliminating the effects of nonlinear interaction on signal accumulation.
Our AIQM scheme achieves better metrological performance than conventional schemes, particularly under strong nonlinear interaction and prolonged signal accumulation, with pronounced robustness against parameter variations. 
By selectively eliminating and utilizing nonlinear interactions via AIQM, our work enables high-precision and high-accuracy entanglement-enhanced sensing without the need for active control of the nonlinear interaction.

\end{abstract}

\maketitle

{\it Introduction. --}
Quantum metrology enables the high-precision measurements essential for various fundamental studies and technological applications~\cite{doi:10.1126/science.1104149,RevModPhys.89.035002,RevModPhys.90.035005,PhysRevLett.132.190001,huang2024entanglementenhanced,MONTENEGRO20251,q6d1-594v}.
For an ensemble of $N$ two-level individual particles, the precision in measuring the energy difference $\hbar\omega_0$ is fundamentally bounded by the standard quantum limit (SQL), $\Delta(\omega_0)_{\rm SQL} \propto 1/(t_s \sqrt{N})$ with $t_s$ denoting the signal accumulation time ~\cite{RevModPhys.90.035005,huang2024entanglementenhanced}.  
Multiparticle quantum entanglement offers a viable path to surpass the SQL, in which the metrologically useful entangled states can be generated via nonlinear interaction such as one-axis twisting (OAT)~\cite{PhysRevA.47.5138,PhysRevA.110.042619}, twist-and-turn~\cite{PhysRevA.67.013607,doi:10.1126/science.1250147,PhysRevA.92.023603,PhysRevA.99.022329,PhysRevA.105.062456}, and two-axis twisting (TAT)~\cite{PhysRevA.92.013623,PhysRevLett.107.013601,PhysRevA.86.012311}.
These protocols can even approach the ultimate precision bound known as the Heisenberg limit, $\Delta(\omega_0)_{\rm HL}\propto 1/(t_s N)$~\cite{RevModPhys.90.035005,huang2024entanglementenhanced}.

A central challenge in entanglement-enhanced quantum metrology is that the nonlinear interactions, necessary for generating metrologically useful entangled states, typically degrade precision if they persist during signal accumulation~\cite{PhysRevX.10.031003,PhysRevX.10.031002}.
Conventional solutions require precise activation, deactivation, and reversal of these interactions—a level of control that remains experimentally demanding~\cite{PhysRevLett.100.140401,gross2010nonlinear,riedel2010atom}.
While dynamical decoupling offers a solution~\cite{PRL.20.180,PhysRevA.108.062602,PhysRevX.14.031017}, it requires strong control pulses, which inherently leads to control-error accumulation~\cite{PhysRevLett.118.133202,Chen_2025}.
{\it Can we effectively suppress nonlinear interaction during signal accumulation without active control of the interaction?}

\begin{figure*}[ht]
    \centering
    \includegraphics[width=1\textwidth]{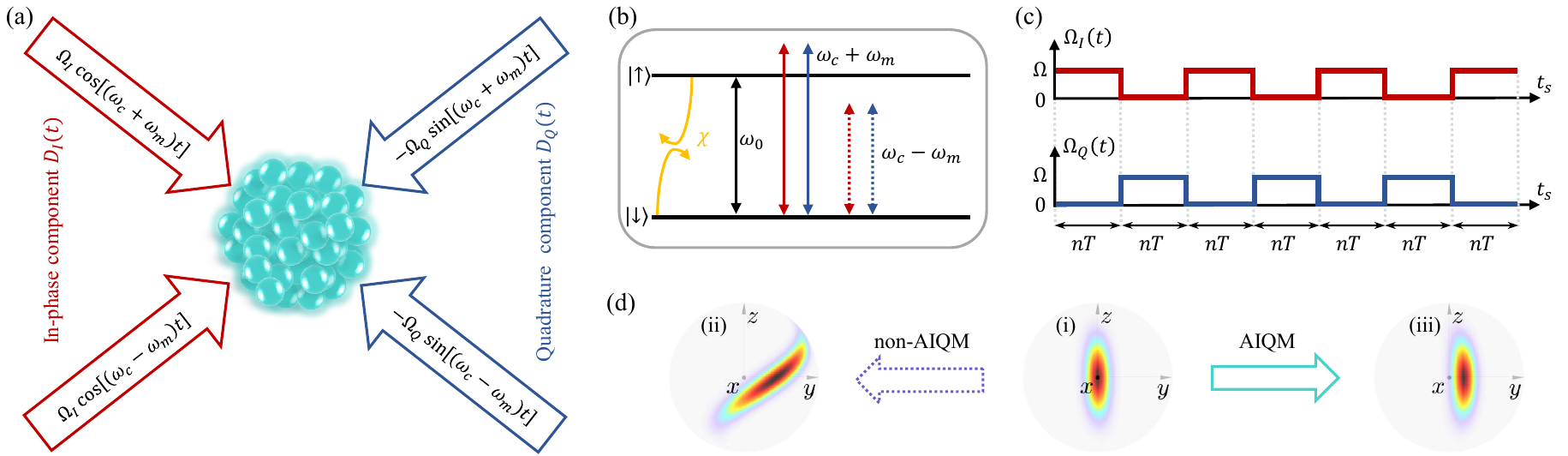} 
    \caption
    {(a) Schematic of an ensemble of interacting two-level particles under in-phase and quadrature driving fields.
    (b) Energy level diagram of the particle. Here, $\chi$ stands for atom-atom interaction strength, $\omega_{0}$ is the transition frequency to be estimated, $\omega_{c}$ is the carrier frequency, and $\omega_{m}$ is the modulation frequency.
    (c) Time sequence of alternating in-phase and quadrature modulation (AIQM) during a signal accumulation duration of $t_s=2mnT$. The in-phase and quadrature Rabi frequencies are periodically alternated between $\left[ \Omega_{I}(t)=\Omega, \Omega_{Q}(t)=0 \right]$ and $\left[\Omega_{I}(t)=0, \Omega_{Q}(t)=\Omega\right]$ at the intervals of $nT=2n\pi/\omega_{m}$. 
    (d) State evolution during signal accumulation with and without AIQM. 
    }
    \label{fig1}
\end{figure*}

Nonlinear interactions are also vital for interaction-based readout.
This technique enables Heisenberg-limited precision without requiring single-particle resolved measurement~\cite{PhysRevLett.115.163002,doi:10.1126/science.aaf3397,PhysRevLett.116.053601,PhysRevA.94.010102,PhysRevA.97.053618,PhysRevLett.119.193601,PhysRevA.98.012129,PhysRevA.98.030303,PhysRevLett.116.090801,PhysRevLett.118.150401,PhysRevA.97.043813,Niezgoda_2019,Schulte2020ramsey,PhysRevResearch.4.013236,PhysRevA.107.032610,mao2023quantum,PhysRevResearch.6.013034,PhysRevResearch.6.033197,PhysRevApplied.22.044066,10.1093/nsr/nwaf091,PhysRevA.111.053709,PhysRevA.97.032116,PhysRevApplied.16.064056,liu2022nonlinear,PhysRevA.107.052613,huang2021dynamic,hu2024nonlinear,PhysRevA.110.022407,gao2025signal,32n5-mhk1} and had been validated in several proof-of-principle experiments~\cite{colombo_time-reversal-based_2022,li2023improving,zaporski2025quantum,PhysRevLett.117.013001,Linnemann_2017,hudelist2014quantum,PhysRevLett.119.223604,Tian:23,garttner2017measuring,doi:10.1126/science.aaw2884,doi:10.1126/science.abi5226}.
However, implementing the full measurement procedure, which includes entangled state generation, signal accumulation, and interaction-based readout, requires precise control over the nonlinear interaction. 
This entails the ability to switch the nonlinear interaction on and off, and even to reverse its sign, a capability that remains experimentally challenging on platforms of Bose-condensed atoms~\cite{PhysRevLett.100.140401,gross2010nonlinear,riedel2010atom}.
{\it Is it possible to achieve Heisenberg-limited precision while keeping the nonlinear interaction fixed throughout the entire procedure?}

In this Letter, we propose an alternating in-phase and quadrature modulation (AIQM) protocol to suppress the OAT interaction during signal accumulation in entanglement-enhanced quantum metrology.
By periodically alternating between in-phase and quadrature driving fields, our protocol suppresses the detrimental effects of the OAT interaction.
In addition to broad applicability, the protocol demonstrates two key advantages: enhanced performance under strong OAT and long accumulation times, and pronounced robustness to parameter variations.
Furthermore, the AIQM can be used for both entanglement preparation and interaction-based readout.
Under a fixed OAT interaction, we present a full-stage AIQM protocol that exhibits significantly better metrological performance than a protocol without AIQM, and demonstrates strong robustness against detection noise.
%
% Under a fixed OAT interaction, we present a full-stage AIQM protocol that exhibits significantly better metrological performance and stronger robustness against detection noise than a protocol without AIQM.
%
Our work introduces an AIQM protocol that advances quantum metrology without requiring control of the nonlinear interaction. 
This opens a door to high-precision, detection-noise-robust sensing in interacting multiparticle quantum systems.

{\it General OAT system under in-phase and quadrature modulations. --}
We consider a general OAT system for $N$ interacting identical particles occupying two distinct energy levels denoted by $\lvert \uparrow\rangle$ and $\lvert \downarrow\rangle$, such as an ensemble of Bose-condensed atoms involving two modes~\cite{PhysRevA.59.620,PhysRevLett.97.150402,lee2012nonlinear,PhysRevA.103.023309}.
Mathematically, the system can be characterized by a collective spin vector $\hat{\bm{J}}=\{\hat{J}_{x},\hat{J}_{y},\hat{J}_{z}\}$ where $\hat{J}_{x}=\frac{1}{2}(\hat{a}^{\dagger}\hat{b}+\hat{b}^{\dagger}\hat{a})$, $\hat{J}_{y}=\frac{1}{2i}(\hat{a}^{\dagger}\hat{b}-\hat{b}^{\dagger}\hat{a})$, and $\hat{J}_{z}=\frac{1}{2}(\hat{a}^{\dagger}\hat{a}-\hat{b}^{\dagger}\hat{b})$~\cite{PhysRevA.33.4033}.
The total spin operator satisfies $\hat{\bm{J}}^2=\hat{J}_{x}^{2}+\hat{J}_{y}^{2}+\hat{J}_{z}^{2}=J\left(J+1\right)$ with $J=N/2$.
Here, $\hat{a}^{\dagger}$ ($\hat{a}$) and $\hat{b}^{\dagger}$ ($\hat{b}$) denote the bosonic creation (annihilation) operators for the particles in $\lvert \uparrow\rangle$ and $\lvert \downarrow\rangle$, respectively.
An arbitrary state can be expressed as $\lvert \psi\rangle=\sum_{m=-J}^{J}C_n\lvert J,m\rangle$, where the Dicke states $\lvert J,m\rangle$ satisfy $\hat{\bm{J}}^2\lvert J,m\rangle=J(J+1)\lvert J,m\rangle$ and $\hat{J}_z\lvert J,m\rangle=m\lvert J,m\rangle$ with $m=-J,-J+1,...,+J$~\cite{RevModPhys.90.035005,PhysRevLett.125.210503}.

As shown in Fig.~\ref{fig1}~(a), we impose an in-phase driving field $D_{I}(t)=\Omega_{I}\cos\left[(\omega_{c}+\omega_{m})t\right]+ \Omega_{I}\cos\left[(\omega_{c}-\omega_{m})t\right]$ and a quadrature driving field $D_{Q}(t)=-\Omega_{Q}\sin\left[(\omega_{c}+\omega_{m})t\right]-\Omega_{Q}\sin\left[(\omega_{c}-\omega_{m})t\right]$ to the OAT system. 
Here $\omega_{c}$ is the carrier frequency, $\omega_{m}$ is the modulation frequency, and $\Omega_{I}$ and $\Omega_{Q}$ are respectively the Rabi frequencies of in-phase and quadrature components.
The overall driving field reads 
\begin{equation}
D_{I}(t)+D_{Q}(t)=V_{I}(t) \cos(\omega_{c} t) - V_{Q}(t) \sin(\omega_{c} t),
\end{equation}
where $V_{I}(t)=2\Omega_{I}\cos(\omega_{m} t)$ and $V_{Q}(t)=2\Omega_{Q}\cos(\omega_{m} t)$.
Therefore the system obeys the Hamiltonian (we set $\hbar=1$ hereafter):
\begin{equation}\label{Ham}
    \hat{H}= \chi \hat{J}_{z}^{2} + \omega_{0} \hat J_z + 2\left[D_{I}(t)+D_{Q}(t)\right]\hat{J}_{x},
\end{equation}
where the transition frequency $\omega_{0}$ between $\lvert \uparrow\rangle$ and $\lvert\downarrow\rangle$ is the physical quantity to be measured, and $\chi$ is the OAT interaction strength, see Fig.~\ref{fig1}~(b).
In the rotating-frame with $\hat{U}_{r}=e^{-i\omega_{c}\hat{J}_{z}}$, under the rotating-wave approximation $\omega_c\gg\left(\Omega, \omega_{m}, N\chi\right)$, the Hamiltonian becomes
\begin{equation}\label{H_sp}
\hat{H}_{R}=\chi\hat{J}_{z}^2+\delta \hat{J}_{z} +2\Omega\cos(\omega_{m} t)(\cos\alpha \hat{J}_{x}+\sin\alpha \hat{J}_{y}),
\end{equation}
where $\delta=\omega_{0}-\omega_{c}$ is the detuning between the transition frequency and the carrier frequency, $\Omega_{I}=\Omega\cos\alpha$, $\Omega_{Q}=\Omega\sin\alpha$, and $\alpha=\arctan(\Omega_{Q}/\Omega_{I})$.

{\it Suppressing OAT interaction via AIQM. --}
Below we illustrate how to suppress the nonlinear OAT interaction via AIQM.
When the modulation frequency is sufficiently large ($\omega_{m} \gg N\chi$), the evolution operator of stroboscopic dynamics at moments ${t}=nT$ ($T=2\pi/\omega_{m}$, $n=1,2,...$) can be described by $\hat{U}_{\rm st}({t}) = \mathcal{\hat{T}} e^{-i \int_{0}^{{t}} \hat{H}_{R}(\tau) d\tau}=e^{-i  \hat{H}_{{s,F}} {t} }$. 
The effective time-independent Floquet Hamiltonian reads $  \hat{H}_{{s,F}} = -\frac{\chi}{2} \left[(1+L_{0})\hat{J}_{\alpha}^{2} + 2 L_{0} \hat{J}_{\beta}^{2} \right] + K_{0} \delta \hat{J}_{z}$ with $\hat{J}_{\alpha} = \cos\alpha \hat{J}_{x} + \sin\alpha \hat{J}_{y}$, $\hat{J}_{\beta} = \cos\beta \hat{J}_{x} + \sin\beta \hat{J}_{y}$, $\beta = \alpha + \pi/2$, $L_{0} = \mathcal{J}_{0}(4 \Omega/\omega_{m})$, and $K_{0} = \mathcal{J}_{0}(2\Omega/\omega_{m})$. 
Here, $\mathcal{J}_{0}$ denotes the zeroth-order Bessel function of the first kind, see more details in Supplemental Material~\cite{SM}.
%
% The derivation of the Floquet Hamiltonian~\eqref{H_sf} is presented in SM.
% 
By tuning $\Omega/\omega_{m}$ such that $L_0 = \mathcal{J}_{0}(4 \Omega/\omega_{m}) = -1/3$, which approximately corresponds to $\Omega/\omega_{m} \simeq 0.8131$, the effective Hamiltonian $\hat{H}_{{s,F}}$ simplifies to $\hat{H}_{{s,F}}^{\rm eff}= \chi_{\rm eff} \left( \hat{J}_{\beta}^{2} - \hat{J}_{\alpha}^{2}\right) + \delta_{\rm eff} \hat{J}_{z}$, 
with the effective interaction strength $\chi_\mathrm{eff} = \chi/3$ and the effective detuning $\delta_\mathrm{eff} = K_0 \delta \simeq 0.4404\,\delta$.
When only in-phase modulation is applied, i.e. $\Omega_{I}=\Omega$ and $\Omega_{Q}=0$, we have $\alpha = 0$ and $\beta = \pi/2$, thus the Hamiltonian reduces to
\begin{equation}
\hat{H}_{s1}^{\rm eff}= \chi_{\rm eff}\left(\hat{J}_{y}^{2} - \hat{J}_{x}^{2}\right) + \delta_{\rm eff} \hat{J}_{z}.
\end{equation}
Conversely, when only quadrature modulation is applied, i.e. $\Omega_{I}=0$ and $\Omega_{Q}=\Omega$, we have $\alpha = \pi/2$ and $\beta = \pi$, thus the Hamiltonian becomes  
\begin{equation}
\hat{H}_{s2}^{\rm eff}= -\chi_{\rm eff}\left(\hat{J}_{y}^{2} - \hat{J}_{x}^{2}\right) + \delta_{\rm eff} \hat{J}_{z}.
\end{equation}

The key to eliminating the OAT interaction lies in the periodic alternation between in-phase and quadrature modulations.
That is, the Rabi frequencies are periodically alternated between $\left[\Omega_{I}(t)=\Omega, \Omega_{Q}(t)=0\right]$ and $\left[\Omega_{I}(t)=0, \Omega_{Q}(t)=\Omega\right]$, see Fig.~\ref{fig1}~(c).
In the first $nT$ duration, setting $\Omega_{I}=\Omega$ and $\Omega_{Q}=0$, the time evolution operator reads
$\hat{U}_{s1}=e^{-i\hat{H}_{s1}^{\rm eff} nT}=e^{-i\left[\chi_{\rm eff}\left( \hat{J}_{y}^{2} - \hat{J}_{x}^{2}\right) + \delta_{\rm eff} \hat{J}_{z}\right] nT}$.
While in the next $nT$ duration, setting $\Omega_{I}=0$ and $\Omega_{Q}=\Omega$, the evolution operator reads
$\hat{U}_{s2}=e^{-i\hat{H}_{s2}^{\rm eff} nT}=e^{-i\left[-\chi_{\rm eff}\left( \hat{J}_{y}^{2} - \hat{J}_{x}^{2}\right) + \delta_{\rm eff} \hat{J}_{z}\right] nT}$.
Combining these two operations, the joint evolution operator for one cycle is $\hat{U}_{s}^{\rm eff}=\hat{U}_{s2}\hat{U}_{s1}=e^{-i\hat{H}_{s2}^{\rm eff} nT}e^{-i\hat{H}_{s1}^{\rm eff} nT}$.
In the limit of sufficiently small $nT$~\cite{Van-Brunt_2015,PhysRevA.90.013604}, 
the effective evolution operator for signal accumulation can be approximated as $\hat{U}_{s}^{\rm eff} (2nT) \approx e^{-i\hat{H}_{s}^{\rm eff} 2nT} =e^{-i \delta_{\mathrm{eff}}\hat{J}_{z} 2nT}$.
This means that the signal accumulation obeys the effective Hamiltonian
\begin{equation}\label{H_eff_s}
    \hat{H}_{s}^{\rm eff}=\delta_{\mathrm{eff}}\hat{J}_{z},
\end{equation}
which completely eliminates the effects of the OAT interaction. 
Thus, for a signal accumulation time of $t_{s} = 2knT$, the overall evolution operator can be given as $\hat{U}_{s}^{\rm eff}(t_{s})=(\hat{U}_{s2}\hat{U}_{s1})^{k}\approx e^{-i\delta_{\mathrm{eff}}\hat{J}_{z} t_s}$.
Consequently, the OAT interaction is effectively suppressed during signal accumulation through AIQM.

\begin{figure}[t]
    \centering
    \includegraphics[width=1\linewidth]{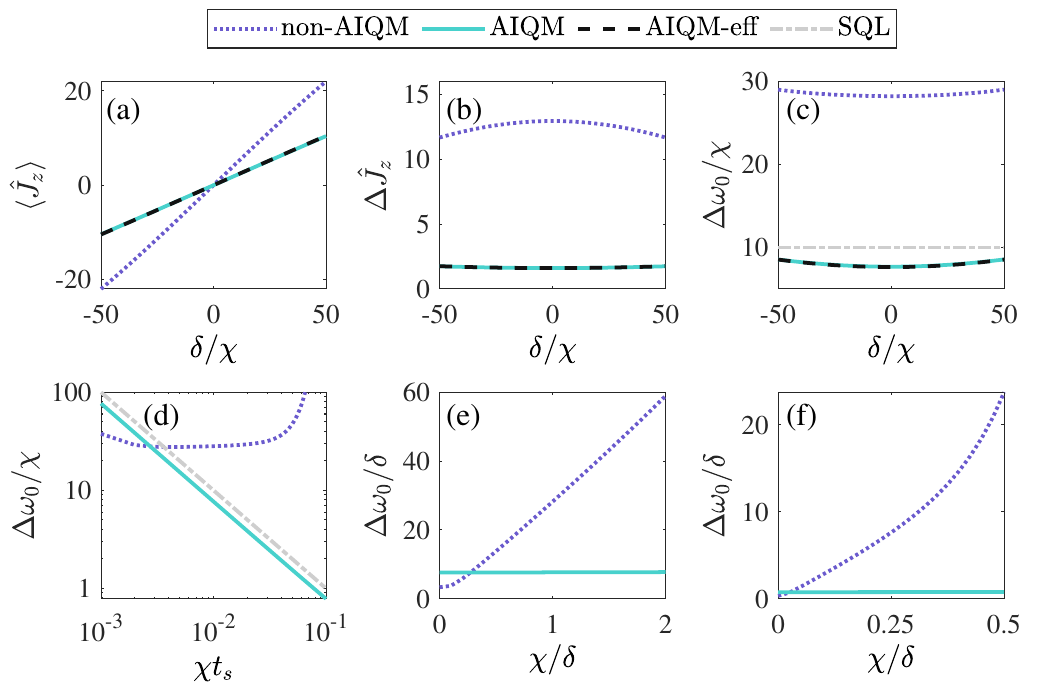}
    \caption{(a) The population imbalance expectation $\langle \hat{J}_z  \rangle$, (b) the population imbalance uncertainty $\Delta\hat{J}_z$, and (c) the estimation precision $\Delta\omega_{0}$ for a OAT system with AIQM (cyan solid line) and without AIQM (purple dotted line). 
    The dependence of $\Delta \omega_{0}$ on (d) the rescaled signal accumulation time $\chi t_s$ and the interaction strength $\chi$ for (e) $t_s = 0.01\delta^{-1}$) and (f) $t_s = 0.1\delta^{-1}$. 
    The black dashed lines are obtained from the effective Hamiltonian~\eqref{H_eff_s}. 
    The gray dash-dotted line indicates the SQL. 
    Here, $N=100$, $\omega_{m}=2\pi\times 20N\chi$, $t_s=0.01\chi^{-1}$, $\chi=1$ for (a)-(d), $\delta=1$ for (d)-(f).}
    \label{fig2}
\end{figure}

The transition frequency $\omega_0$ can be estimated by measuring the population imbalance $\hat{J}_z$.
To show the entanglement enhancement, the input state is chosen as a spin-squeezed state $\lvert\psi_{p}\rangle=e^{-i \hat{J}_x (\pi-2\gamma)}e^{-i \chi \hat{J}_{z}^{2} t_{en}}\lvert\psi_{x}\rangle$ generated by OAT dynamics from a spin coherent state $\lvert\psi_{x}\rangle$, see (i) in Fig.~\ref{fig1}~(d).
Here $\chi t_{en}=0.03$ and $\gamma=1/2\arctan\left(B/A\right)$ with ${A}=1-\cos^{N-2}\left(2\chi t_{en}\right)$ and ${B}=4\cos^{N-2}\left(\chi t_{en}\right)\sin\left(\chi t_{en}\right)$~\cite{MA201189}. 
Without AIQM, the state becomes $\lvert\psi_{1}\rangle=e^{-i (\chi \hat{J}_{z}^{2}+\delta \hat{J}_z) t_s}\lvert\psi_{p}\rangle$ with amplified variance of $\hat{J}_{y}$, see (ii) in Fig.~\ref{fig1}~(d).
In contrast, with AIQM, the state becomes $\lvert\psi_{2}\rangle\approx e^{-i\delta_{\mathrm{eff}}\hat{J}_{z} t_s}\lvert\psi_{p}\rangle$, which only rotates along $\hat{J}_z$ and the variance of $\hat{J}_{y}$ is not amplified, see (iii) in Fig.~\ref{fig1}~(d).
To measure the population imbalance $\hat J_z$, we apply a rotation $\hat{R}_{x}^{\pi/2}$ for readout. 
Thus the estimation precision can be given by $\Delta \omega_{0}={ \Delta \hat{J}_z}/{\vert \partial {\langle \hat{J}_z  \rangle} /\partial {\omega_{0}} \vert}$ with $\Delta \hat{J}_z =\sqrt{ \langle \hat{J}_z^{2}  \rangle-{ \langle \hat{J}_z  \rangle}^{2}}$, $ \langle \hat{J}_z \rangle= \langle \psi_{f} \rvert\hat{J}_z \lvert \psi_{f} \rangle$ and $ \langle \hat{J}_z^2 \rangle= \langle \psi_{f} \rvert\hat{J}_z^2 \lvert \psi_{f} \rangle$.
To verify the effective Hamiltonian~\eqref{H_eff_s},  
we calculate $\langle \hat{J}_z  \rangle$, $\Delta \hat{J}_z$, and  $\Delta \omega_{0}$ for the schemes with and without AIQM, see Fig.~\ref{fig2}.
Although the signal response $\partial { \langle \hat{J}_z  \rangle}/\partial {\omega_{0}}$ for the AIQM protocol is slightly lower [see Fig.~\ref{fig2}~(a)], the standard deviation $\Delta \hat J_z$ is strongly suppressed [see Fig.~\ref{fig2}~(b)]. 
This yields a high precision $\Delta \omega_{0}$ below the SQL, as shown in Fig.~\ref{fig2}~(c).
Our numerical results agree well with analytical predictions based on $\hat{H}_{s}^{\rm eff}$, confirming that AIQM can efficiently suppress the OAT interaction.

In Fig.~\ref{fig2}~(d), we show the dependence of the estimation precision $\Delta \omega_{0} $ on the rescaled signal accumulation time $\chi t_s$.  
Without AIQM, the estimation precision initially decreases with $\chi t_s$ for short accumulation times but subsequently increases as the detrimental effects of the OAT interaction become significant.
This degrades estimation performance worse than the SQL. 
In contrast, with AIQM, as $\chi t_s$ increases, the estimation precision decreases monotonically with $ \Delta \omega_{0} \propto 1/(\chi t_s)$ and keeps better than the SQL.
Thus, our AIQM approach may utilize the signal accumulation time as a scalable resource to enhance measurement sensitivity by efficiently suppressing the OAT interaction.

To assess the impact of the OAT strength $ \chi $, we examine the estimation precision $\Delta \omega_0$ for different signal accumulation times: $ t = 0.01\delta^{-1} $ [see Fig.~\ref{fig2}~(e)] and $ t = 0.1\delta^{-1} $ [see Fig.~\ref{fig2}~(f)].
In the absence of AIQM, $\Delta \omega_0$ increases with $\chi$, reflecting the adverse influence of stronger nonlinear OAT interaction.  
In contrast, $\Delta \omega_0$ remains unchanged under AIQM across all $\chi$, demonstrating robustness even in regimes of strong interaction.  
Moreover, as shown in Fig.~\ref{fig2}~(f), even for weak OAT interaction, long accumulation time can lead to significant precision degradation in the absence of AIQM, whereas the AIQM scheme maintains high estimation precision.
This highlights the broad applicability and advantage of our AIQM method, even under strong OAT interactions and long accumulation times.

\begin{figure}[t]
    \centering
    \includegraphics[width=1\linewidth]{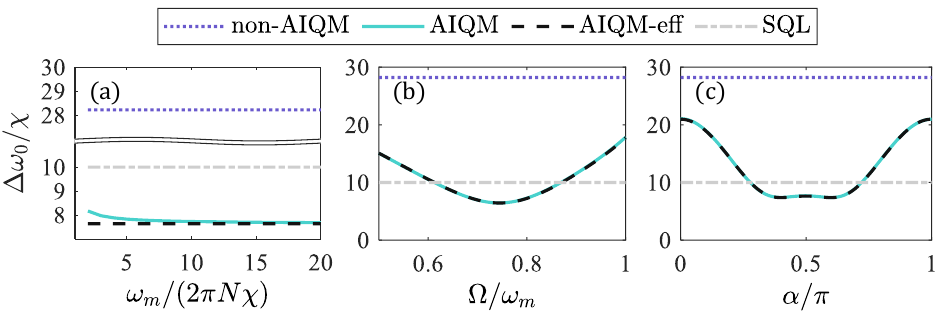}
    \caption{The robustness of estimation precision $\Delta \omega_{0}$ against different control parameters: (a) the modulation frequency $\omega_{m}$, (b) the ratio of Rabi frequencies $\Omega/\omega_{m}$, and (c) the phase $\alpha=\arctan(\Omega_{Q}/\Omega_{I})$ in the second period. Here, $N=100$,  $t_s=0.01\chi^{-1}$, $\omega_{m}=2\pi\times 20N\chi$, $\delta=\chi=1$.
    % The results of theoretical prediction are denoted as black dashed line. The gray dash-dotted line indicates the SQL. 
    }
    \label{fig3}
\end{figure}

{\it Parameter robustness. --}
Our AIQM protocol exhibits strong robustness against control parameters.
First, under the condition of $\mathcal{J}_{0}(4 \Omega/\omega_{m}) = -1/3$, we examine the robustness against the modulation frequency $\omega_m$.  
As $\omega_m$ increases, the rescaled estimation precision $\Delta \omega_0/\chi$ given by the original time-dependent Hamiltonian~\eqref{H_sp} (the cyan solid line) converge rapidly toward the one (the black dashed line) obtained from the effective time-independent Hamiltonian~\eqref{H_eff_s}.
Obviously, the estimation precision under AIQM is below the SQL (the gray dashed line), see Fig.~\ref{fig3}~(a). 
Notably, when $\omega_m/(2\pi N \chi)>5$, the AIQM scheme efficiently suppresses the OAT interaction and significantly enhances the estimation precision.
Second, we analyze the robustness against the ratio $\Omega/\omega_{m}$, see Fig.~\ref{fig3}~(b).
The $\Delta \omega_0/\chi$ given by the original Hamiltonian~\eqref{H_sp} (the cyan solid line) agrees well with the one (the black dashed line) obtained from the effective Hamiltonian $\hat{H}_{\rm eff,1} = -\frac{\chi}{4} \left[(1+3L_{0})\hat{J}_{x}^{2} + (1+3L_{0})\hat{J}_{y}^{2}\right]+ K_{0} \delta \hat{J}_{z}$ with $L_{0} = \mathcal{J}_{0}(4 \Omega/\omega_{m})$ and $K_{0} = \mathcal{J}_{0}(2\Omega/\omega_{m})$.
Notably, over the interval $\Omega/\omega_{m}\in[0.62, 0.87]$, the estimation precision under AIQM is better than the SQL.
Third, we assess the robustness against the phase $\alpha$ in the second period, while keeping it as zero in the first period, see Fig.~\ref{fig3}~(c). 
Thus the effective Hamiltonian reads $\hat{H}_{\rm eff,2} =\chi_{\rm eff}\left[ (\cos{\alpha})^2(\hat{J}_{y}^{2}-\hat{J}_{x}^{2})-\frac{1}{2}\sin{2\alpha}(\hat{J}_{x}\hat{J}_{y}+\hat{J}_{y}\hat{J}_{x})\right] + \delta_{\mathrm{eff}} \hat{J}_{z}$. 
Over the interval $\alpha/\pi\in[0.28, 0.72]$, the estimation precision under AIQM is better than the SQL.
In particular, for all three cases, the estimation precisions under AIQM are always better than the ones without AIQM (the purple dotted lines).

\begin{figure}[t]
    \centering
    \includegraphics[width=1\linewidth]{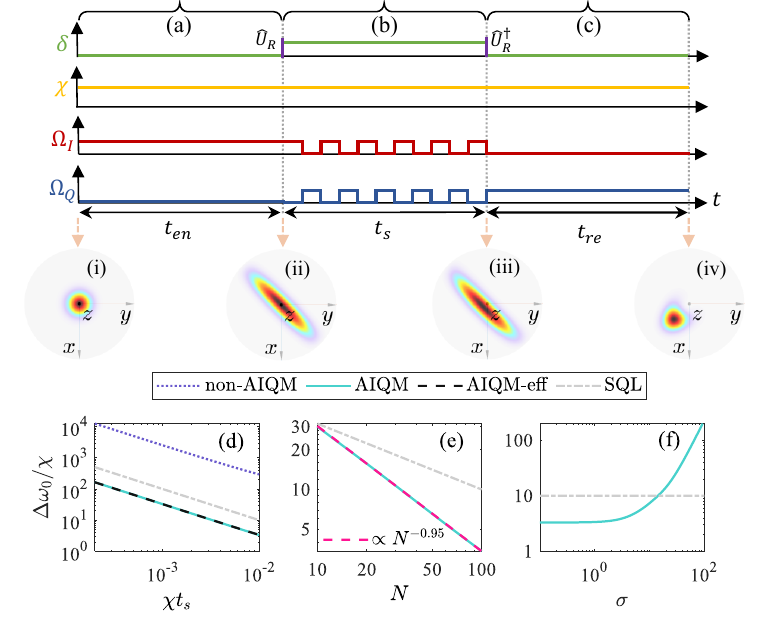} 
    \caption
    {The full-stage AIQM protocol for quantum metrology via selective utilization and suppression of OAT interaction. Under a fixed OAT interaction, the AIQM is applied in the whole procedure: (a) entanglement preparation, (b) signal accumulation, and (c) interaction-based readout. 
    (d) The performance of schemes with (the cyan solid line) and without (the purple dotted line) AIQM. The result of effective time-independent Hamiltonians (the black dashed line) is shown for comparison.
    (e) The precision scaling versus the particle number $N$. The pink dashed line is the fitted for the AIQM protocol with $\Delta \omega_{0} /\chi=270 N^{-0.95}$.
    (f) The robustness against the detection noise for $t_s=0.01\chi^{-1}$. Here, $N=100$, $\omega_{m}=2\pi\times 100N\chi$, and $\delta=\chi=1$.
    }
    \label{fig4}
\end{figure}

{\it Utilizing and suppressing OAT in full-stage AIQM protocol. --}
The in-phase and quadrature modulations can also be used for entanglement preparation and interaction-based readout. 
To prepare the multiparticle entanglement, setting $\Omega_{I}=\Omega$, $\Omega_{Q}=0$ and $\delta=0$, an effective TAT $\hat{H}_{en}^{\rm eff}=\chi_{\rm eff}\left(\hat{J}_{y}^{2} - \hat{J}_{x}^{2}\right)$ can be obtained from the Hamiltonian~\eqref{H_sp}~\cite{SM}, enabling preparation of spin-squeezed states with squeezing parameter scaling as $N^{-1}$~\cite{PhysRevA.90.013604,PhysRevA.91.043642}.
To implement the interaction-based readout, setting $\Omega_{I}=0$, $\Omega_{Q}=\Omega$ and $\delta=0$, a corresponding time-reversal TAT $\hat{H}_{re}^{\rm eff}=-\chi_{\rm eff}\left(\hat{J}_{y}^{2} - \hat{J}_{x}^{2}\right)$ can be obtained~\cite{SM}.
Thus combing the entanglement preparation, the signal accumulation and the interaction-based readout via AIQM, we present a full-stage AIQM protocol in the presence of OAT interaction, see Fig.~\ref{fig4}~(a)-(c).

Now we analyze the state evolution in the full-stage AIQM protocol.
Initially, a spin-coherent state along $z$-axis $\lvert\psi_{z}\rangle$ is prepared, see Fig.~\ref{fig4}~(i). 
Then, setting $\Omega_{I}=0$, $\Omega_{Q}=\Omega$ and $\delta=0$, a spin-squeezed state $\lvert\psi_{en}\rangle \approx e^{-i\hat{H}_{en}^{\rm eff} t_{en}}\lvert\psi_{z}\rangle$ is generated, see Fig.~\ref{fig4}~(ii).
Prior to signal accumulation, a suitable rotation $\hat{U}_{R} = e^{-i\hat{J}_{x}\pi/4} e^{i\hat{J}_{y}\pi/2}$ is applied to prepare the desired entangled state $\lvert\psi_{en}'\rangle = \hat{U}_{R}\lvert\psi_{en}\rangle$ along $x$-axis, which has minimal squeezing along $y$-axis.
Then the signal is accumulated according to the effective Hamiltonian~\eqref{H_eff_s}, that is, the state becomes $\lvert\psi_{s}\rangle \approx e^{-i\hat{H}_{s}^{\rm eff} t_{s}} \lvert\psi_{en}'\rangle$ after signal accumulation. 
To rotate $\lvert\psi_{s}\rangle$ back into $xy$-plane, an additional rotation $\hat{U}_{R}^{\dagger} = e^{-i\hat{J}_{y}\pi/2} e^{i\hat{J}_{x}\pi/4}$ is applied and the state becomes $\lvert\psi_{s}'\rangle=\hat{U}_{R}^{\dagger}\lvert\psi_{s}\rangle$. 
As shown in Fig.~\ref{fig4}~(iii), the transformation $\hat{U}_{R}^{\dagger} e^{-i\hat{H}_{s}^{\rm eff} t_{s}} \hat{U}_{R} = e^{-i \delta_{\rm eff} \hat{J}_\gamma t_s}$ is equivalent to encoding the signal along $\hat{J}_\gamma = \frac{\sqrt{2}}{2}(\hat{J}_x + \hat{J}_y)$ from $\lvert\psi_{en}\rangle$.
In the interaction-based readout stage, setting $\Omega_{I}=0$, $\Omega_{Q}=\Omega$ and $\delta=0$, the effective time-reversal TAT recovers the entangled state into a nearly spin-coherent state $\lvert\psi_{re}\rangle \approx e^{-i\hat{H}_{re}^{\rm eff} t_{re}} \lvert\psi_{s}'\rangle$.
This leads to a significant displacement along $\hat{J}_{\gamma^{'}}=\frac{\sqrt{2}}{2}(\hat{J}_x-\hat{J}_y)$, see Fig.~\ref{fig4}~(iv).
Finally, another rotation $\hat{U}_{R'}=e^{i \hat{J}_x 3\pi/4}e^{-i \hat{J}_y \pi/2} $ is applied, and the population imbalance $\hat J_z$ of the final state $\lvert\psi_{f}\rangle=\hat{U}_{R'}\lvert\psi_{re}\rangle$ is measured.

To show the validity of the effective model for the full-stage AIQM protocol, we compare its results with those of the original time-dependent Hamiltonian. The preparation and readout times are set to $t_{en}=t_{re}=3\ln(2N)/(2N) \chi^{-1}$, corresponding to optimal performance~\cite{PhysRevA.110.022407,PhysRevA.92.013623}.
As shown in Fig.~\ref{fig4}~(d), the results from the effective and original Hamiltonians overlap perfectly and beat the SQL. 
In contrast, the protocol without AIQM fails to suppress OAT during signal accumulation, disables interaction‑based readout, and performs significantly worse.
We also examine the scaling with respect to the particle number $N$, see Fig.~\ref{fig4}~(e). 
Our AIQM protocol consistently surpasses the SQL, with a fitted relation $\Delta \omega_{0} /\chi=270 N^{-0.95}$, approaching the Heisenberg scaling $1/N$. 
In addition, the estimation precision $\Delta \omega_{0}$ versus the detection-noise strength $\sigma$~\cite{SM} is shown in Fig.~\ref{fig4}~(f).
When $\sigma$ is below $\sqrt{N}$, $\Delta \omega_{0}$ remains better than the SQL, demonstrating strong robustness against the detection noise.

{\it Conclusion and discussion. --}
In summary, we propose an AIQM protocol to suppress the detrimental nonlinear interaction during the signal accumulation in entanglement-enhanced quantum metrology.
The detrimental effects of the OAT interaction is efficiently suppressed via employing periodic alternation between in-phase and quadrature driving fields.
The scheme enhances metrological performance under strong OAT interaction and long accumulation time and is robust against control parameter.
In further, we propose a full-stage AIQM protocol, where the OAT interaction keeps fixed throughout the whole procedure. 
Compared to the schemes without AIQM, our protocol exhibits superior precision and strong robustness against detection noise. 
Therefore our AIQM protocol enables high‑precision, detection‑noise‑resilient metrology without controlling the nonlinear interaction.
Beyond improving quantum metrology, the periodic modulations developed in our AIQM can be used to achieve versatile Floquet engineering in quantum manipulation~\cite{annurev}. 
Moreover, the time-reversal dynamics accessible through AIQM offer a valuable tool for probing out-of-time-order correlations, quantum chaos, and information scrambling in serval experimental platforms where direct inversion of nonlinear interactions is infeasible~\cite{PhysRevA.94.040302,PhysRevX.7.031011,PhysRevLett.120.040402,lewis2019unifying,landsman2019verified,PhysRevLett.125.240605,PhysRevLett.124.240505,braumuller2022probing,li2025error}.

Our AIQM protocol can be implemented using various experimental systems of OAT interaction, such as, Bose-condensed atoms~\cite{gross2010nonlinear,riedel2010atom}, cavity-QED~\cite{PhysRevLett.104.073602,greve2022entanglement}, trapped-ions~\cite{bohnet2016quantum}, Rydberg-dressed atomic gas~\cite{PhysRevLett.131.063401}, hot atomic ensemble~\cite{xkt1-y58b}, and nuclear magnetic resonance~\cite{PhysRevLett.114.043604}.
In-phase and quadrature modulations, the key components of our protocol, have been well developed and widely used in lots of experiments.
For instance, using Bose-condensed atoms, our analysis shows that all required parameters are within current experimental capabilities~\cite{SM}.
This strongly supports the near-term experimental realization of the AIQM scheme.

{\it Acknowledgments. --}
The authors thank Yi Shen, Sijie Chen, and Chengyin Han for helpful discussions. This work is supported by the National Natural Science Foundation of China (Grants No.~92476201, No.~12475029, and No.~12025509), the National Key Research and Development Program of China (Grant No. 2022YFA1404104), and the Guangdong Provincial Quantum Science Strategic Initiative (GDZX2305006 and GDZX2405002).

\nocite{}


\begin{thebibliography}{95}%
\makeatletter
\providecommand \@ifxundefined [1]{%
 \@ifx{#1\undefined}
}%
\providecommand \@ifnum [1]{%
 \ifnum #1\expandafter \@firstoftwo
 \else \expandafter \@secondoftwo
 \fi
}%
\providecommand \@ifx [1]{%
 \ifx #1\expandafter \@firstoftwo
 \else \expandafter \@secondoftwo
 \fi
}%
\providecommand \natexlab [1]{#1}%
\providecommand \enquote  [1]{``#1''}%
\providecommand \bibnamefont  [1]{#1}%
\providecommand \bibfnamefont [1]{#1}%
\providecommand \citenamefont [1]{#1}%
\providecommand \href@noop [0]{\@secondoftwo}%
\providecommand \href [0]{\begingroup \@sanitize@url \@href}%
\providecommand \@href[1]{\@@startlink{#1}\@@href}%
\providecommand \@@href[1]{\endgroup#1\@@endlink}%
\providecommand \@sanitize@url [0]{\catcode `\\12\catcode `\$12\catcode `\&12\catcode `\#12\catcode `\^12\catcode `\_12\catcode `\%12\relax}%
\providecommand \@@startlink[1]{}%
\providecommand \@@endlink[0]{}%
\providecommand \url  [0]{\begingroup\@sanitize@url \@url }%
\providecommand \@url [1]{\endgroup\@href {#1}{\urlprefix }}%
\providecommand \urlprefix  [0]{URL }%
\providecommand \Eprint [0]{\href }%
\providecommand \doibase [0]{https://doi.org/}%
\providecommand \selectlanguage [0]{\@gobble}%
\providecommand \bibinfo  [0]{\@secondoftwo}%
\providecommand \bibfield  [0]{\@secondoftwo}%
\providecommand \translation [1]{[#1]}%
\providecommand \BibitemOpen [0]{}%
\providecommand \bibitemStop [0]{}%
\providecommand \bibitemNoStop [0]{.\EOS\space}%
\providecommand \EOS [0]{\spacefactor3000\relax}%
\providecommand \BibitemShut  [1]{\csname bibitem#1\endcsname}%
\let\auto@bib@innerbib\@empty
%</preamble>
\bibitem [{\citenamefont {Giovannetti}\ \emph {et~al.}(2004)\citenamefont {Giovannetti}, \citenamefont {Lloyd},\ and\ \citenamefont {Maccone}}]{doi:10.1126/science.1104149}%
  \BibitemOpen
  \bibfield  {author} {\bibinfo {author} {\bibfnamefont {V.}~\bibnamefont {Giovannetti}}, \bibinfo {author} {\bibfnamefont {S.}~\bibnamefont {Lloyd}},\ and\ \bibinfo {author} {\bibfnamefont {L.}~\bibnamefont {Maccone}},\ }\bibfield  {title} {\bibinfo {title} {Quantum-enhanced measurements: Beating the standard quantum limit},\ }\href {https://doi.org/10.1126/science.1104149} {\bibfield  {journal} {\bibinfo  {journal} {Science}\ }\textbf {\bibinfo {volume} {306}},\ \bibinfo {pages} {1330} (\bibinfo {year} {2004})}\BibitemShut {NoStop}%
\bibitem [{\citenamefont {Degen}\ \emph {et~al.}(2017)\citenamefont {Degen}, \citenamefont {Reinhard},\ and\ \citenamefont {Cappellaro}}]{RevModPhys.89.035002}%
  \BibitemOpen
  \bibfield  {author} {\bibinfo {author} {\bibfnamefont {C.~L.}\ \bibnamefont {Degen}}, \bibinfo {author} {\bibfnamefont {F.}~\bibnamefont {Reinhard}},\ and\ \bibinfo {author} {\bibfnamefont {P.}~\bibnamefont {Cappellaro}},\ }\bibfield  {title} {\bibinfo {title} {Quantum sensing},\ }\href {https://doi.org/10.1103/RevModPhys.89.035002} {\bibfield  {journal} {\bibinfo  {journal} {Rev. Mod. Phys.}\ }\textbf {\bibinfo {volume} {89}},\ \bibinfo {pages} {035002} (\bibinfo {year} {2017})}\BibitemShut {NoStop}%
\bibitem [{\citenamefont {Pezz\`e}\ \emph {et~al.}(2018)\citenamefont {Pezz\`e}, \citenamefont {Smerzi}, \citenamefont {Oberthaler}, \citenamefont {Schmied},\ and\ \citenamefont {Treutlein}}]{RevModPhys.90.035005}%
  \BibitemOpen
  \bibfield  {author} {\bibinfo {author} {\bibfnamefont {L.}~\bibnamefont {Pezz\`e}}, \bibinfo {author} {\bibfnamefont {A.}~\bibnamefont {Smerzi}}, \bibinfo {author} {\bibfnamefont {M.~K.}\ \bibnamefont {Oberthaler}}, \bibinfo {author} {\bibfnamefont {R.}~\bibnamefont {Schmied}},\ and\ \bibinfo {author} {\bibfnamefont {P.}~\bibnamefont {Treutlein}},\ }\bibfield  {title} {\bibinfo {title} {Quantum metrology with nonclassical states of atomic ensembles},\ }\href {https://doi.org/10.1103/RevModPhys.90.035005} {\bibfield  {journal} {\bibinfo  {journal} {Rev. Mod. Phys.}\ }\textbf {\bibinfo {volume} {90}},\ \bibinfo {pages} {035005} (\bibinfo {year} {2018})}\BibitemShut {NoStop}%
\bibitem [{\citenamefont {Ye}\ and\ \citenamefont {Zoller}(2024)}]{PhysRevLett.132.190001}%
  \BibitemOpen
  \bibfield  {author} {\bibinfo {author} {\bibfnamefont {J.}~\bibnamefont {Ye}}\ and\ \bibinfo {author} {\bibfnamefont {P.}~\bibnamefont {Zoller}},\ }\bibfield  {title} {\bibinfo {title} {Essay: Quantum sensing with atomic, molecular, and optical platforms for fundamental physics},\ }\href {https://doi.org/10.1103/PhysRevLett.132.190001} {\bibfield  {journal} {\bibinfo  {journal} {Phys. Rev. Lett.}\ }\textbf {\bibinfo {volume} {132}},\ \bibinfo {pages} {190001} (\bibinfo {year} {2024})}\BibitemShut {NoStop}%
\bibitem [{\citenamefont {Huang}\ \emph {et~al.}(2024)\citenamefont {Huang}, \citenamefont {Zhuang},\ and\ \citenamefont {Lee}}]{huang2024entanglementenhanced}%
  \BibitemOpen
  \bibfield  {author} {\bibinfo {author} {\bibfnamefont {J.}~\bibnamefont {Huang}}, \bibinfo {author} {\bibfnamefont {M.}~\bibnamefont {Zhuang}},\ and\ \bibinfo {author} {\bibfnamefont {C.}~\bibnamefont {Lee}},\ }\bibfield  {title} {\bibinfo {title} {{Entanglement-enhanced quantum metrology: From standard quantum limit to {Heisenberg} limit}},\ }\href {https://doi.org/10.1063/5.0204102} {\bibfield  {journal} {\bibinfo  {journal} {Appl. Phys. Rev.}\ }\textbf {\bibinfo {volume} {11}},\ \bibinfo {pages} {031302} (\bibinfo {year} {2024})}\BibitemShut {NoStop}%
\bibitem [{\citenamefont {Montenegro}\ \emph {et~al.}(2025)\citenamefont {Montenegro}, \citenamefont {Mukhopadhyay}, \citenamefont {Yousefjani}, \citenamefont {Sarkar}, \citenamefont {Mishra}, \citenamefont {Paris},\ and\ \citenamefont {Bayat}}]{MONTENEGRO20251}%
  \BibitemOpen
  \bibfield  {author} {\bibinfo {author} {\bibfnamefont {V.}~\bibnamefont {Montenegro}}, \bibinfo {author} {\bibfnamefont {C.}~\bibnamefont {Mukhopadhyay}}, \bibinfo {author} {\bibfnamefont {R.}~\bibnamefont {Yousefjani}}, \bibinfo {author} {\bibfnamefont {S.}~\bibnamefont {Sarkar}}, \bibinfo {author} {\bibfnamefont {U.}~\bibnamefont {Mishra}}, \bibinfo {author} {\bibfnamefont {M.~G.}\ \bibnamefont {Paris}},\ and\ \bibinfo {author} {\bibfnamefont {A.}~\bibnamefont {Bayat}},\ }\bibfield  {title} {\bibinfo {title} {Review: Quantum metrology and sensing with many-body systems},\ }\href {https://doi.org/https://doi.org/10.1016/j.physrep.2025.05.005} {\bibfield  {journal} {\bibinfo  {journal} {Physics Reports}\ }\textbf {\bibinfo {volume} {1134}},\ \bibinfo {pages} {1} (\bibinfo {year} {2025})}\BibitemShut {NoStop}%
\bibitem [{\citenamefont {Brown}\ and\ \citenamefont {Walker}(2025)}]{q6d1-594v}%
  \BibitemOpen
  \bibfield  {author} {\bibinfo {author} {\bibfnamefont {J.~M.}\ \bibnamefont {Brown}}\ and\ \bibinfo {author} {\bibfnamefont {T.~G.}\ \bibnamefont {Walker}},\ }\bibfield  {title} {\bibinfo {title} {Perspective: Practical atom-based quantum sensors},\ }\href {https://doi.org/10.1103/q6d1-594v} {\bibfield  {journal} {\bibinfo  {journal} {Phys. Rev. A}\ }\textbf {\bibinfo {volume} {112}},\ \bibinfo {pages} {040102} (\bibinfo {year} {2025})}\BibitemShut {NoStop}%
\bibitem [{\citenamefont {Kitagawa}\ and\ \citenamefont {Ueda}(1993)}]{PhysRevA.47.5138}%
  \BibitemOpen
  \bibfield  {author} {\bibinfo {author} {\bibfnamefont {M.}~\bibnamefont {Kitagawa}}\ and\ \bibinfo {author} {\bibfnamefont {M.}~\bibnamefont {Ueda}},\ }\bibfield  {title} {\bibinfo {title} {Squeezed spin states},\ }\href {https://doi.org/10.1103/PhysRevA.47.5138} {\bibfield  {journal} {\bibinfo  {journal} {Phys. Rev. A}\ }\textbf {\bibinfo {volume} {47}},\ \bibinfo {pages} {5138} (\bibinfo {year} {1993})}\BibitemShut {NoStop}%
\bibitem [{\citenamefont {Shen}\ \emph {et~al.}(2024)\citenamefont {Shen}, \citenamefont {Zhou}, \citenamefont {Huang},\ and\ \citenamefont {Lee}}]{PhysRevA.110.042619}%
  \BibitemOpen
  \bibfield  {author} {\bibinfo {author} {\bibfnamefont {Y.}~\bibnamefont {Shen}}, \bibinfo {author} {\bibfnamefont {J.}~\bibnamefont {Zhou}}, \bibinfo {author} {\bibfnamefont {J.}~\bibnamefont {Huang}},\ and\ \bibinfo {author} {\bibfnamefont {C.}~\bibnamefont {Lee}},\ }\bibfield  {title} {\bibinfo {title} {Quantum metrology with partially concurrent twisting and sensing},\ }\href {https://doi.org/10.1103/PhysRevA.110.042619} {\bibfield  {journal} {\bibinfo  {journal} {Phys. Rev. A}\ }\textbf {\bibinfo {volume} {110}},\ \bibinfo {pages} {042619} (\bibinfo {year} {2024})}\BibitemShut {NoStop}%
\bibitem [{\citenamefont {Micheli}\ \emph {et~al.}(2003)\citenamefont {Micheli}, \citenamefont {Jaksch}, \citenamefont {Cirac},\ and\ \citenamefont {Zoller}}]{PhysRevA.67.013607}%
  \BibitemOpen
  \bibfield  {author} {\bibinfo {author} {\bibfnamefont {A.}~\bibnamefont {Micheli}}, \bibinfo {author} {\bibfnamefont {D.}~\bibnamefont {Jaksch}}, \bibinfo {author} {\bibfnamefont {J.~I.}\ \bibnamefont {Cirac}},\ and\ \bibinfo {author} {\bibfnamefont {P.}~\bibnamefont {Zoller}},\ }\bibfield  {title} {\bibinfo {title} {Many-particle entanglement in two-component {Bose-Einstein} condensates},\ }\href {https://doi.org/10.1103/PhysRevA.67.013607} {\bibfield  {journal} {\bibinfo  {journal} {Phys. Rev. A}\ }\textbf {\bibinfo {volume} {67}},\ \bibinfo {pages} {013607} (\bibinfo {year} {2003})}\BibitemShut {NoStop}%
\bibitem [{\citenamefont {Strobel}\ \emph {et~al.}(2014)\citenamefont {Strobel}, \citenamefont {Muessel}, \citenamefont {Linnemann}, \citenamefont {Zibold}, \citenamefont {Hume}, \citenamefont {Pezzè}, \citenamefont {Smerzi},\ and\ \citenamefont {Oberthaler}}]{doi:10.1126/science.1250147}%
  \BibitemOpen
  \bibfield  {author} {\bibinfo {author} {\bibfnamefont {H.}~\bibnamefont {Strobel}}, \bibinfo {author} {\bibfnamefont {W.}~\bibnamefont {Muessel}}, \bibinfo {author} {\bibfnamefont {D.}~\bibnamefont {Linnemann}}, \bibinfo {author} {\bibfnamefont {T.}~\bibnamefont {Zibold}}, \bibinfo {author} {\bibfnamefont {D.~B.}\ \bibnamefont {Hume}}, \bibinfo {author} {\bibfnamefont {L.}~\bibnamefont {Pezzè}}, \bibinfo {author} {\bibfnamefont {A.}~\bibnamefont {Smerzi}},\ and\ \bibinfo {author} {\bibfnamefont {M.~K.}\ \bibnamefont {Oberthaler}},\ }\bibfield  {title} {\bibinfo {title} {Fisher information and entanglement of {non-Gaussian} spin states},\ }\href {https://doi.org/10.1126/science.1250147} {\bibfield  {journal} {\bibinfo  {journal} {Science}\ }\textbf {\bibinfo {volume} {345}},\ \bibinfo {pages} {424} (\bibinfo {year} {2014})}\BibitemShut {NoStop}%
\bibitem [{\citenamefont {Muessel}\ \emph {et~al.}(2015)\citenamefont {Muessel}, \citenamefont {Strobel}, \citenamefont {Linnemann}, \citenamefont {Zibold}, \citenamefont {Juli\'a-D\'{\i}az},\ and\ \citenamefont {Oberthaler}}]{PhysRevA.92.023603}%
  \BibitemOpen
  \bibfield  {author} {\bibinfo {author} {\bibfnamefont {W.}~\bibnamefont {Muessel}}, \bibinfo {author} {\bibfnamefont {H.}~\bibnamefont {Strobel}}, \bibinfo {author} {\bibfnamefont {D.}~\bibnamefont {Linnemann}}, \bibinfo {author} {\bibfnamefont {T.}~\bibnamefont {Zibold}}, \bibinfo {author} {\bibfnamefont {B.}~\bibnamefont {Juli\'a-D\'{\i}az}},\ and\ \bibinfo {author} {\bibfnamefont {M.~K.}\ \bibnamefont {Oberthaler}},\ }\bibfield  {title} {\bibinfo {title} {Twist-and-turn spin squeezing in {Bose-Einstein} condensates},\ }\href {https://doi.org/10.1103/PhysRevA.92.023603} {\bibfield  {journal} {\bibinfo  {journal} {Phys. Rev. A}\ }\textbf {\bibinfo {volume} {92}},\ \bibinfo {pages} {023603} (\bibinfo {year} {2015})}\BibitemShut {NoStop}%
\bibitem [{\citenamefont {Sorelli}\ \emph {et~al.}(2019)\citenamefont {Sorelli}, \citenamefont {Gessner}, \citenamefont {Smerzi},\ and\ \citenamefont {Pezz\`e}}]{PhysRevA.99.022329}%
  \BibitemOpen
  \bibfield  {author} {\bibinfo {author} {\bibfnamefont {G.}~\bibnamefont {Sorelli}}, \bibinfo {author} {\bibfnamefont {M.}~\bibnamefont {Gessner}}, \bibinfo {author} {\bibfnamefont {A.}~\bibnamefont {Smerzi}},\ and\ \bibinfo {author} {\bibfnamefont {L.}~\bibnamefont {Pezz\`e}},\ }\bibfield  {title} {\bibinfo {title} {Fast and optimal generation of entanglement in bosonic {Josephson} junctions},\ }\href {https://doi.org/10.1103/PhysRevA.99.022329} {\bibfield  {journal} {\bibinfo  {journal} {Phys. Rev. A}\ }\textbf {\bibinfo {volume} {99}},\ \bibinfo {pages} {022329} (\bibinfo {year} {2019})}\BibitemShut {NoStop}%
\bibitem [{\citenamefont {Huang}\ \emph {et~al.}(2022)\citenamefont {Huang}, \citenamefont {Huo}, \citenamefont {Zhuang},\ and\ \citenamefont {Lee}}]{PhysRevA.105.062456}%
  \BibitemOpen
  \bibfield  {author} {\bibinfo {author} {\bibfnamefont {J.}~\bibnamefont {Huang}}, \bibinfo {author} {\bibfnamefont {H.}~\bibnamefont {Huo}}, \bibinfo {author} {\bibfnamefont {M.}~\bibnamefont {Zhuang}},\ and\ \bibinfo {author} {\bibfnamefont {C.}~\bibnamefont {Lee}},\ }\bibfield  {title} {\bibinfo {title} {Efficient generation of spin cat states with twist-and-turn dynamics via machine optimization},\ }\href {https://doi.org/10.1103/PhysRevA.105.062456} {\bibfield  {journal} {\bibinfo  {journal} {Phys. Rev. A}\ }\textbf {\bibinfo {volume} {105}},\ \bibinfo {pages} {062456} (\bibinfo {year} {2022})}\BibitemShut {NoStop}%
\bibitem [{\citenamefont {Kajtoch}\ and\ \citenamefont {Witkowska}(2015)}]{PhysRevA.92.013623}%
  \BibitemOpen
  \bibfield  {author} {\bibinfo {author} {\bibfnamefont {D.}~\bibnamefont {Kajtoch}}\ and\ \bibinfo {author} {\bibfnamefont {E.}~\bibnamefont {Witkowska}},\ }\bibfield  {title} {\bibinfo {title} {Quantum dynamics generated by the two-axis countertwisting {Hamiltonian}},\ }\href {https://doi.org/10.1103/PhysRevA.92.013623} {\bibfield  {journal} {\bibinfo  {journal} {Phys. Rev. A}\ }\textbf {\bibinfo {volume} {92}},\ \bibinfo {pages} {013623} (\bibinfo {year} {2015})}\BibitemShut {NoStop}%
\bibitem [{\citenamefont {Liu}\ \emph {et~al.}(2011)\citenamefont {Liu}, \citenamefont {Xu}, \citenamefont {Jin},\ and\ \citenamefont {You}}]{PhysRevLett.107.013601}%
  \BibitemOpen
  \bibfield  {author} {\bibinfo {author} {\bibfnamefont {Y.~C.}\ \bibnamefont {Liu}}, \bibinfo {author} {\bibfnamefont {Z.~F.}\ \bibnamefont {Xu}}, \bibinfo {author} {\bibfnamefont {G.~R.}\ \bibnamefont {Jin}},\ and\ \bibinfo {author} {\bibfnamefont {L.}~\bibnamefont {You}},\ }\bibfield  {title} {\bibinfo {title} {Spin squeezing: Transforming one-axis twisting into two-axis twisting},\ }\href {https://doi.org/10.1103/PhysRevLett.107.013601} {\bibfield  {journal} {\bibinfo  {journal} {Phys. Rev. Lett.}\ }\textbf {\bibinfo {volume} {107}},\ \bibinfo {pages} {013601} (\bibinfo {year} {2011})}\BibitemShut {NoStop}%
\bibitem [{\citenamefont {Chaudhry}\ and\ \citenamefont {Gong}(2012)}]{PhysRevA.86.012311}%
  \BibitemOpen
  \bibfield  {author} {\bibinfo {author} {\bibfnamefont {A.~Z.}\ \bibnamefont {Chaudhry}}\ and\ \bibinfo {author} {\bibfnamefont {J.}~\bibnamefont {Gong}},\ }\bibfield  {title} {\bibinfo {title} {Protecting and enhancing spin squeezing via continuous dynamical decoupling},\ }\href {https://doi.org/10.1103/PhysRevA.86.012311} {\bibfield  {journal} {\bibinfo  {journal} {Phys. Rev. A}\ }\textbf {\bibinfo {volume} {86}},\ \bibinfo {pages} {012311} (\bibinfo {year} {2012})}\BibitemShut {NoStop}%
\bibitem [{\citenamefont {Zhou}\ \emph {et~al.}(2020)\citenamefont {Zhou}, \citenamefont {Choi}, \citenamefont {Choi}, \citenamefont {Landig}, \citenamefont {Douglas}, \citenamefont {Isoya}, \citenamefont {Jelezko}, \citenamefont {Onoda}, \citenamefont {Sumiya}, \citenamefont {Cappellaro}, \citenamefont {Knowles}, \citenamefont {Park},\ and\ \citenamefont {Lukin}}]{PhysRevX.10.031003}%
  \BibitemOpen
  \bibfield  {author} {\bibinfo {author} {\bibfnamefont {H.}~\bibnamefont {Zhou}}, \bibinfo {author} {\bibfnamefont {J.}~\bibnamefont {Choi}}, \bibinfo {author} {\bibfnamefont {S.}~\bibnamefont {Choi}}, \bibinfo {author} {\bibfnamefont {R.}~\bibnamefont {Landig}}, \bibinfo {author} {\bibfnamefont {A.~M.}\ \bibnamefont {Douglas}}, \bibinfo {author} {\bibfnamefont {J.}~\bibnamefont {Isoya}}, \bibinfo {author} {\bibfnamefont {F.}~\bibnamefont {Jelezko}}, \bibinfo {author} {\bibfnamefont {S.}~\bibnamefont {Onoda}}, \bibinfo {author} {\bibfnamefont {H.}~\bibnamefont {Sumiya}}, \bibinfo {author} {\bibfnamefont {P.}~\bibnamefont {Cappellaro}}, \bibinfo {author} {\bibfnamefont {H.~S.}\ \bibnamefont {Knowles}}, \bibinfo {author} {\bibfnamefont {H.}~\bibnamefont {Park}},\ and\ \bibinfo {author} {\bibfnamefont {M.~D.}\ \bibnamefont {Lukin}},\ }\bibfield  {title} {\bibinfo {title} {Quantum metrology with strongly interacting spin systems},\ }\href {https://doi.org/10.1103/PhysRevX.10.031003} {\bibfield  {journal}
  {\bibinfo  {journal} {Phys. Rev. X}\ }\textbf {\bibinfo {volume} {10}},\ \bibinfo {pages} {031003} (\bibinfo {year} {2020})}\BibitemShut {NoStop}%
\bibitem [{\citenamefont {Choi}\ \emph {et~al.}(2020)\citenamefont {Choi}, \citenamefont {Zhou}, \citenamefont {Knowles}, \citenamefont {Landig}, \citenamefont {Choi},\ and\ \citenamefont {Lukin}}]{PhysRevX.10.031002}%
  \BibitemOpen
  \bibfield  {author} {\bibinfo {author} {\bibfnamefont {J.}~\bibnamefont {Choi}}, \bibinfo {author} {\bibfnamefont {H.}~\bibnamefont {Zhou}}, \bibinfo {author} {\bibfnamefont {H.~S.}\ \bibnamefont {Knowles}}, \bibinfo {author} {\bibfnamefont {R.}~\bibnamefont {Landig}}, \bibinfo {author} {\bibfnamefont {S.}~\bibnamefont {Choi}},\ and\ \bibinfo {author} {\bibfnamefont {M.~D.}\ \bibnamefont {Lukin}},\ }\bibfield  {title} {\bibinfo {title} {Robust dynamic {Hamiltonian} engineering of many-body spin systems},\ }\href {https://doi.org/10.1103/PhysRevX.10.031002} {\bibfield  {journal} {\bibinfo  {journal} {Phys. Rev. X}\ }\textbf {\bibinfo {volume} {10}},\ \bibinfo {pages} {031002} (\bibinfo {year} {2020})}\BibitemShut {NoStop}%
\bibitem [{\citenamefont {Widera}\ \emph {et~al.}(2008)\citenamefont {Widera}, \citenamefont {Trotzky}, \citenamefont {Cheinet}, \citenamefont {F\"olling}, \citenamefont {Gerbier}, \citenamefont {Bloch}, \citenamefont {Gritsev}, \citenamefont {Lukin},\ and\ \citenamefont {Demler}}]{PhysRevLett.100.140401}%
  \BibitemOpen
  \bibfield  {author} {\bibinfo {author} {\bibfnamefont {A.}~\bibnamefont {Widera}}, \bibinfo {author} {\bibfnamefont {S.}~\bibnamefont {Trotzky}}, \bibinfo {author} {\bibfnamefont {P.}~\bibnamefont {Cheinet}}, \bibinfo {author} {\bibfnamefont {S.}~\bibnamefont {F\"olling}}, \bibinfo {author} {\bibfnamefont {F.}~\bibnamefont {Gerbier}}, \bibinfo {author} {\bibfnamefont {I.}~\bibnamefont {Bloch}}, \bibinfo {author} {\bibfnamefont {V.}~\bibnamefont {Gritsev}}, \bibinfo {author} {\bibfnamefont {M.~D.}\ \bibnamefont {Lukin}},\ and\ \bibinfo {author} {\bibfnamefont {E.}~\bibnamefont {Demler}},\ }\bibfield  {title} {\bibinfo {title} {Quantum spin dynamics of mode-squeezed {Luttinger} liquids in two-component atomic gases},\ }\href {https://doi.org/10.1103/PhysRevLett.100.140401} {\bibfield  {journal} {\bibinfo  {journal} {Phys. Rev. Lett.}\ }\textbf {\bibinfo {volume} {100}},\ \bibinfo {pages} {140401} (\bibinfo {year} {2008})}\BibitemShut {NoStop}%
\bibitem [{\citenamefont {Gross}\ \emph {et~al.}(2010)\citenamefont {Gross}, \citenamefont {Zibold}, \citenamefont {Nicklas}, \citenamefont {Esteve},\ and\ \citenamefont {Oberthaler}}]{gross2010nonlinear}%
  \BibitemOpen
  \bibfield  {author} {\bibinfo {author} {\bibfnamefont {C.}~\bibnamefont {Gross}}, \bibinfo {author} {\bibfnamefont {T.}~\bibnamefont {Zibold}}, \bibinfo {author} {\bibfnamefont {E.}~\bibnamefont {Nicklas}}, \bibinfo {author} {\bibfnamefont {J.}~\bibnamefont {Esteve}},\ and\ \bibinfo {author} {\bibfnamefont {M.~K.}\ \bibnamefont {Oberthaler}},\ }\bibfield  {title} {\bibinfo {title} {Nonlinear atom interferometer surpasses classical precision limit},\ }\href {https://doi.org/10.1038/nature08919} {\bibfield  {journal} {\bibinfo  {journal} {Nature}\ }\textbf {\bibinfo {volume} {464}},\ \bibinfo {pages} {1165} (\bibinfo {year} {2010})}\BibitemShut {NoStop}%
\bibitem [{\citenamefont {Riedel}\ \emph {et~al.}(2010)\citenamefont {Riedel}, \citenamefont {B{\"o}hi}, \citenamefont {Li}, \citenamefont {H{\"a}nsch}, \citenamefont {Sinatra},\ and\ \citenamefont {Treutlein}}]{riedel2010atom}%
  \BibitemOpen
  \bibfield  {author} {\bibinfo {author} {\bibfnamefont {M.~F.}\ \bibnamefont {Riedel}}, \bibinfo {author} {\bibfnamefont {P.}~\bibnamefont {B{\"o}hi}}, \bibinfo {author} {\bibfnamefont {Y.}~\bibnamefont {Li}}, \bibinfo {author} {\bibfnamefont {T.~W.}\ \bibnamefont {H{\"a}nsch}}, \bibinfo {author} {\bibfnamefont {A.}~\bibnamefont {Sinatra}},\ and\ \bibinfo {author} {\bibfnamefont {P.}~\bibnamefont {Treutlein}},\ }\bibfield  {title} {\bibinfo {title} {Atom-chip-based generation of entanglement for quantum metrology},\ }\href {https://doi.org/10.1038/nature08988} {\bibfield  {journal} {\bibinfo  {journal} {Nature}\ }\textbf {\bibinfo {volume} {464}},\ \bibinfo {pages} {1170} (\bibinfo {year} {2010})}\BibitemShut {NoStop}%
\bibitem [{\citenamefont {Waugh}\ \emph {et~al.}(1968)\citenamefont {Waugh}, \citenamefont {Huber},\ and\ \citenamefont {Haeberlen}}]{PRL.20.180}%
  \BibitemOpen
  \bibfield  {author} {\bibinfo {author} {\bibfnamefont {J.~S.}\ \bibnamefont {Waugh}}, \bibinfo {author} {\bibfnamefont {L.~M.}\ \bibnamefont {Huber}},\ and\ \bibinfo {author} {\bibfnamefont {U.}~\bibnamefont {Haeberlen}},\ }\bibfield  {title} {\bibinfo {title} {Approach to high-resolution nmr in solids},\ }\href {https://doi.org/10.1103/PhysRevLett.20.180} {\bibfield  {journal} {\bibinfo  {journal} {Phys. Rev. Lett.}\ }\textbf {\bibinfo {volume} {20}},\ \bibinfo {pages} {180} (\bibinfo {year} {1968})}\BibitemShut {NoStop}%
\bibitem [{\citenamefont {Tyler}\ \emph {et~al.}(2023)\citenamefont {Tyler}, \citenamefont {Zhou}, \citenamefont {Martin}, \citenamefont {Leitao},\ and\ \citenamefont {Lukin}}]{PhysRevA.108.062602}%
  \BibitemOpen
  \bibfield  {author} {\bibinfo {author} {\bibfnamefont {M.}~\bibnamefont {Tyler}}, \bibinfo {author} {\bibfnamefont {H.}~\bibnamefont {Zhou}}, \bibinfo {author} {\bibfnamefont {L.~S.}\ \bibnamefont {Martin}}, \bibinfo {author} {\bibfnamefont {N.}~\bibnamefont {Leitao}},\ and\ \bibinfo {author} {\bibfnamefont {M.~D.}\ \bibnamefont {Lukin}},\ }\bibfield  {title} {\bibinfo {title} {Higher-order methods for {Hamiltonian} engineering pulse sequence design},\ }\href {https://doi.org/10.1103/PhysRevA.108.062602} {\bibfield  {journal} {\bibinfo  {journal} {Phys. Rev. A}\ }\textbf {\bibinfo {volume} {108}},\ \bibinfo {pages} {062602} (\bibinfo {year} {2023})}\BibitemShut {NoStop}%
\bibitem [{\citenamefont {Zhou}\ \emph {et~al.}(2024{\natexlab{a}})\citenamefont {Zhou}, \citenamefont {Gao}, \citenamefont {Leitao}, \citenamefont {Makarova}, \citenamefont {Cong}, \citenamefont {Douglas}, \citenamefont {Martin},\ and\ \citenamefont {Lukin}}]{PhysRevX.14.031017}%
  \BibitemOpen
  \bibfield  {author} {\bibinfo {author} {\bibfnamefont {H.}~\bibnamefont {Zhou}}, \bibinfo {author} {\bibfnamefont {H.}~\bibnamefont {Gao}}, \bibinfo {author} {\bibfnamefont {N.~T.}\ \bibnamefont {Leitao}}, \bibinfo {author} {\bibfnamefont {O.}~\bibnamefont {Makarova}}, \bibinfo {author} {\bibfnamefont {I.}~\bibnamefont {Cong}}, \bibinfo {author} {\bibfnamefont {A.~M.}\ \bibnamefont {Douglas}}, \bibinfo {author} {\bibfnamefont {L.~S.}\ \bibnamefont {Martin}},\ and\ \bibinfo {author} {\bibfnamefont {M.~D.}\ \bibnamefont {Lukin}},\ }\bibfield  {title} {\bibinfo {title} {Robust {Hamiltonian} engineering for interacting qudit systems},\ }\href {https://doi.org/10.1103/PhysRevX.14.031017} {\bibfield  {journal} {\bibinfo  {journal} {Phys. Rev. X}\ }\textbf {\bibinfo {volume} {14}},\ \bibinfo {pages} {031017} (\bibinfo {year} {2024}{\natexlab{a}})}\BibitemShut {NoStop}%
\bibitem [{\citenamefont {Genov}\ \emph {et~al.}(2017)\citenamefont {Genov}, \citenamefont {Schraft}, \citenamefont {Vitanov},\ and\ \citenamefont {Halfmann}}]{PhysRevLett.118.133202}%
  \BibitemOpen
  \bibfield  {author} {\bibinfo {author} {\bibfnamefont {G.~T.}\ \bibnamefont {Genov}}, \bibinfo {author} {\bibfnamefont {D.}~\bibnamefont {Schraft}}, \bibinfo {author} {\bibfnamefont {N.~V.}\ \bibnamefont {Vitanov}},\ and\ \bibinfo {author} {\bibfnamefont {T.}~\bibnamefont {Halfmann}},\ }\bibfield  {title} {\bibinfo {title} {Arbitrarily accurate pulse sequences for robust dynamical decoupling},\ }\href {https://doi.org/10.1103/PhysRevLett.118.133202} {\bibfield  {journal} {\bibinfo  {journal} {Phys. Rev. Lett.}\ }\textbf {\bibinfo {volume} {118}},\ \bibinfo {pages} {133202} (\bibinfo {year} {2017})}\BibitemShut {NoStop}%
\bibitem [{\citenamefont {Chen}\ \emph {et~al.}(2025{\natexlab{a}})\citenamefont {Chen}, \citenamefont {Chen}, \citenamefont {Huang}, \citenamefont {Liu}, \citenamefont {Zhuang},\ and\ \citenamefont {Lee}}]{Chen_2025}%
  \BibitemOpen
  \bibfield  {author} {\bibinfo {author} {\bibfnamefont {S.}~\bibnamefont {Chen}}, \bibinfo {author} {\bibfnamefont {G.}~\bibnamefont {Chen}}, \bibinfo {author} {\bibfnamefont {J.}~\bibnamefont {Huang}}, \bibinfo {author} {\bibfnamefont {P.}~\bibnamefont {Liu}}, \bibinfo {author} {\bibfnamefont {M.}~\bibnamefont {Zhuang}},\ and\ \bibinfo {author} {\bibfnamefont {C.}~\bibnamefont {Lee}},\ }\bibfield  {title} {\bibinfo {title} {Phase-modulated dynamical decoupling sequences robust to systematic amplitude error},\ }\href {https://doi.org/10.1088/1674-1056/adcd43} {\bibfield  {journal} {\bibinfo  {journal} {Chinese Physics B}\ }\textbf {\bibinfo {volume} {34}},\ \bibinfo {pages} {074202} (\bibinfo {year} {2025}{\natexlab{a}})}\BibitemShut {NoStop}%
\bibitem [{\citenamefont {Gabbrielli}\ \emph {et~al.}(2015)\citenamefont {Gabbrielli}, \citenamefont {Pezz\`e},\ and\ \citenamefont {Smerzi}}]{PhysRevLett.115.163002}%
  \BibitemOpen
  \bibfield  {author} {\bibinfo {author} {\bibfnamefont {M.}~\bibnamefont {Gabbrielli}}, \bibinfo {author} {\bibfnamefont {L.}~\bibnamefont {Pezz\`e}},\ and\ \bibinfo {author} {\bibfnamefont {A.}~\bibnamefont {Smerzi}},\ }\bibfield  {title} {\bibinfo {title} {Spin-mixing interferometry with {Bose-Einstein} condensates},\ }\href {https://doi.org/10.1103/PhysRevLett.115.163002} {\bibfield  {journal} {\bibinfo  {journal} {Phys. Rev. Lett.}\ }\textbf {\bibinfo {volume} {115}},\ \bibinfo {pages} {163002} (\bibinfo {year} {2015})}\BibitemShut {NoStop}%
\bibitem [{\citenamefont {Hosten}\ \emph {et~al.}(2016)\citenamefont {Hosten}, \citenamefont {Krishnakumar}, \citenamefont {Engelsen},\ and\ \citenamefont {Kasevich}}]{doi:10.1126/science.aaf3397}%
  \BibitemOpen
  \bibfield  {author} {\bibinfo {author} {\bibfnamefont {O.}~\bibnamefont {Hosten}}, \bibinfo {author} {\bibfnamefont {R.}~\bibnamefont {Krishnakumar}}, \bibinfo {author} {\bibfnamefont {N.~J.}\ \bibnamefont {Engelsen}},\ and\ \bibinfo {author} {\bibfnamefont {M.~A.}\ \bibnamefont {Kasevich}},\ }\bibfield  {title} {\bibinfo {title} {Quantum phase magnification},\ }\href {https://doi.org/10.1126/science.aaf3397} {\bibfield  {journal} {\bibinfo  {journal} {Science}\ }\textbf {\bibinfo {volume} {352}},\ \bibinfo {pages} {1552} (\bibinfo {year} {2016})}\BibitemShut {NoStop}%
\bibitem [{\citenamefont {Davis}\ \emph {et~al.}(2016)\citenamefont {Davis}, \citenamefont {Bentsen},\ and\ \citenamefont {Schleier-Smith}}]{PhysRevLett.116.053601}%
  \BibitemOpen
  \bibfield  {author} {\bibinfo {author} {\bibfnamefont {E.}~\bibnamefont {Davis}}, \bibinfo {author} {\bibfnamefont {G.}~\bibnamefont {Bentsen}},\ and\ \bibinfo {author} {\bibfnamefont {M.}~\bibnamefont {Schleier-Smith}},\ }\bibfield  {title} {\bibinfo {title} {Approaching the {Heisenberg} limit without single-particle detection},\ }\href {https://doi.org/10.1103/PhysRevLett.116.053601} {\bibfield  {journal} {\bibinfo  {journal} {Phys. Rev. Lett.}\ }\textbf {\bibinfo {volume} {116}},\ \bibinfo {pages} {053601} (\bibinfo {year} {2016})}\BibitemShut {NoStop}%
\bibitem [{\citenamefont {Macr\`{\i}}\ \emph {et~al.}(2016)\citenamefont {Macr\`{\i}}, \citenamefont {Smerzi},\ and\ \citenamefont {Pezz\`e}}]{PhysRevA.94.010102}%
  \BibitemOpen
  \bibfield  {author} {\bibinfo {author} {\bibfnamefont {T.}~\bibnamefont {Macr\`{\i}}}, \bibinfo {author} {\bibfnamefont {A.}~\bibnamefont {Smerzi}},\ and\ \bibinfo {author} {\bibfnamefont {L.}~\bibnamefont {Pezz\`e}},\ }\bibfield  {title} {\bibinfo {title} {Loschmidt echo for quantum metrology},\ }\href {https://doi.org/10.1103/PhysRevA.94.010102} {\bibfield  {journal} {\bibinfo  {journal} {Phys. Rev. A}\ }\textbf {\bibinfo {volume} {94}},\ \bibinfo {pages} {010102} (\bibinfo {year} {2016})}\BibitemShut {NoStop}%
\bibitem [{\citenamefont {Mirkhalaf}\ \emph {et~al.}(2018)\citenamefont {Mirkhalaf}, \citenamefont {Nolan},\ and\ \citenamefont {Haine}}]{PhysRevA.97.053618}%
  \BibitemOpen
  \bibfield  {author} {\bibinfo {author} {\bibfnamefont {S.~S.}\ \bibnamefont {Mirkhalaf}}, \bibinfo {author} {\bibfnamefont {S.~P.}\ \bibnamefont {Nolan}},\ and\ \bibinfo {author} {\bibfnamefont {S.~A.}\ \bibnamefont {Haine}},\ }\bibfield  {title} {\bibinfo {title} {Robustifying twist-and-turn entanglement with interaction-based readout},\ }\href {https://doi.org/10.1103/PhysRevA.97.053618} {\bibfield  {journal} {\bibinfo  {journal} {Phys. Rev. A}\ }\textbf {\bibinfo {volume} {97}},\ \bibinfo {pages} {053618} (\bibinfo {year} {2018})}\BibitemShut {NoStop}%
\bibitem [{\citenamefont {Nolan}\ \emph {et~al.}(2017)\citenamefont {Nolan}, \citenamefont {Szigeti},\ and\ \citenamefont {Haine}}]{PhysRevLett.119.193601}%
  \BibitemOpen
  \bibfield  {author} {\bibinfo {author} {\bibfnamefont {S.~P.}\ \bibnamefont {Nolan}}, \bibinfo {author} {\bibfnamefont {S.~S.}\ \bibnamefont {Szigeti}},\ and\ \bibinfo {author} {\bibfnamefont {S.~A.}\ \bibnamefont {Haine}},\ }\bibfield  {title} {\bibinfo {title} {Optimal and robust quantum metrology using interaction-based readouts},\ }\href {https://doi.org/10.1103/PhysRevLett.119.193601} {\bibfield  {journal} {\bibinfo  {journal} {Phys. Rev. Lett.}\ }\textbf {\bibinfo {volume} {119}},\ \bibinfo {pages} {193601} (\bibinfo {year} {2017})}\BibitemShut {NoStop}%
\bibitem [{\citenamefont {Huang}\ \emph {et~al.}(2018{\natexlab{a}})\citenamefont {Huang}, \citenamefont {Zhuang}, \citenamefont {Lu}, \citenamefont {Ke},\ and\ \citenamefont {Lee}}]{PhysRevA.98.012129}%
  \BibitemOpen
  \bibfield  {author} {\bibinfo {author} {\bibfnamefont {J.}~\bibnamefont {Huang}}, \bibinfo {author} {\bibfnamefont {M.}~\bibnamefont {Zhuang}}, \bibinfo {author} {\bibfnamefont {B.}~\bibnamefont {Lu}}, \bibinfo {author} {\bibfnamefont {Y.}~\bibnamefont {Ke}},\ and\ \bibinfo {author} {\bibfnamefont {C.}~\bibnamefont {Lee}},\ }\bibfield  {title} {\bibinfo {title} {Achieving {Heisenberg-limited} metrology with spin cat states via interaction-based readout},\ }\href {https://doi.org/10.1103/PhysRevA.98.012129} {\bibfield  {journal} {\bibinfo  {journal} {Phys. Rev. A}\ }\textbf {\bibinfo {volume} {98}},\ \bibinfo {pages} {012129} (\bibinfo {year} {2018}{\natexlab{a}})}\BibitemShut {NoStop}%
\bibitem [{\citenamefont {Haine}(2018)}]{PhysRevA.98.030303}%
  \BibitemOpen
  \bibfield  {author} {\bibinfo {author} {\bibfnamefont {S.~A.}\ \bibnamefont {Haine}},\ }\bibfield  {title} {\bibinfo {title} {Using interaction-based readouts to approach the ultimate limit of detection-noise robustness for quantum-enhanced metrology in collective spin systems},\ }\href {https://doi.org/10.1103/PhysRevA.98.030303} {\bibfield  {journal} {\bibinfo  {journal} {Phys. Rev. A}\ }\textbf {\bibinfo {volume} {98}},\ \bibinfo {pages} {030303} (\bibinfo {year} {2018})}\BibitemShut {NoStop}%
\bibitem [{\citenamefont {Fr\"owis}\ \emph {et~al.}(2016)\citenamefont {Fr\"owis}, \citenamefont {Sekatski},\ and\ \citenamefont {D\"ur}}]{PhysRevLett.116.090801}%
  \BibitemOpen
  \bibfield  {author} {\bibinfo {author} {\bibfnamefont {F.}~\bibnamefont {Fr\"owis}}, \bibinfo {author} {\bibfnamefont {P.}~\bibnamefont {Sekatski}},\ and\ \bibinfo {author} {\bibfnamefont {W.}~\bibnamefont {D\"ur}},\ }\bibfield  {title} {\bibinfo {title} {Detecting large quantum {Fisher} information with finite measurement precision},\ }\href {https://doi.org/10.1103/PhysRevLett.116.090801} {\bibfield  {journal} {\bibinfo  {journal} {Phys. Rev. Lett.}\ }\textbf {\bibinfo {volume} {116}},\ \bibinfo {pages} {090801} (\bibinfo {year} {2016})}\BibitemShut {NoStop}%
\bibitem [{\citenamefont {Szigeti}\ \emph {et~al.}(2017)\citenamefont {Szigeti}, \citenamefont {Lewis-Swan},\ and\ \citenamefont {Haine}}]{PhysRevLett.118.150401}%
  \BibitemOpen
  \bibfield  {author} {\bibinfo {author} {\bibfnamefont {S.~S.}\ \bibnamefont {Szigeti}}, \bibinfo {author} {\bibfnamefont {R.~J.}\ \bibnamefont {Lewis-Swan}},\ and\ \bibinfo {author} {\bibfnamefont {S.~A.}\ \bibnamefont {Haine}},\ }\bibfield  {title} {\bibinfo {title} {Pumped-up {SU(1,1)} interferometry},\ }\href {https://doi.org/10.1103/PhysRevLett.118.150401} {\bibfield  {journal} {\bibinfo  {journal} {Phys. Rev. Lett.}\ }\textbf {\bibinfo {volume} {118}},\ \bibinfo {pages} {150401} (\bibinfo {year} {2017})}\BibitemShut {NoStop}%
\bibitem [{\citenamefont {Anders}\ \emph {et~al.}(2018)\citenamefont {Anders}, \citenamefont {Pezz\`e}, \citenamefont {Smerzi},\ and\ \citenamefont {Klempt}}]{PhysRevA.97.043813}%
  \BibitemOpen
  \bibfield  {author} {\bibinfo {author} {\bibfnamefont {F.}~\bibnamefont {Anders}}, \bibinfo {author} {\bibfnamefont {L.}~\bibnamefont {Pezz\`e}}, \bibinfo {author} {\bibfnamefont {A.}~\bibnamefont {Smerzi}},\ and\ \bibinfo {author} {\bibfnamefont {C.}~\bibnamefont {Klempt}},\ }\bibfield  {title} {\bibinfo {title} {Phase magnification by two-axis countertwisting for detection-noise robust interferometry},\ }\href {https://doi.org/10.1103/PhysRevA.97.043813} {\bibfield  {journal} {\bibinfo  {journal} {Phys. Rev. A}\ }\textbf {\bibinfo {volume} {97}},\ \bibinfo {pages} {043813} (\bibinfo {year} {2018})}\BibitemShut {NoStop}%
\bibitem [{\citenamefont {Niezgoda}\ \emph {et~al.}(2019)\citenamefont {Niezgoda}, \citenamefont {Kajtoch}, \citenamefont {Dziekańska},\ and\ \citenamefont {Witkowska}}]{Niezgoda_2019}%
  \BibitemOpen
  \bibfield  {author} {\bibinfo {author} {\bibfnamefont {A.}~\bibnamefont {Niezgoda}}, \bibinfo {author} {\bibfnamefont {D.}~\bibnamefont {Kajtoch}}, \bibinfo {author} {\bibfnamefont {J.}~\bibnamefont {Dziekańska}},\ and\ \bibinfo {author} {\bibfnamefont {E.}~\bibnamefont {Witkowska}},\ }\bibfield  {title} {\bibinfo {title} {Optimal quantum interferometry robust to detection noise using spin-1 atomic condensates},\ }\href {https://doi.org/10.1088/1367-2630/ab4099} {\bibfield  {journal} {\bibinfo  {journal} {New Journal of Physics}\ }\textbf {\bibinfo {volume} {21}},\ \bibinfo {pages} {093037} (\bibinfo {year} {2019})}\BibitemShut {NoStop}%
\bibitem [{\citenamefont {Schulte}\ \emph {et~al.}(2020)\citenamefont {Schulte}, \citenamefont {Mart{\'{i}}nez-Lahuerta}, \citenamefont {Scharnagl},\ and\ \citenamefont {Hammerer}}]{Schulte2020ramsey}%
  \BibitemOpen
  \bibfield  {author} {\bibinfo {author} {\bibfnamefont {M.}~\bibnamefont {Schulte}}, \bibinfo {author} {\bibfnamefont {V.~J.}\ \bibnamefont {Mart{\'{i}}nez-Lahuerta}}, \bibinfo {author} {\bibfnamefont {M.~S.}\ \bibnamefont {Scharnagl}},\ and\ \bibinfo {author} {\bibfnamefont {K.}~\bibnamefont {Hammerer}},\ }\bibfield  {title} {\bibinfo {title} {Ramsey interferometry with generalized one-axis twisting echoes},\ }\href {https://doi.org/10.22331/q-2020-05-15-268} {\bibfield  {journal} {\bibinfo  {journal} {{Quantum}}\ }\textbf {\bibinfo {volume} {4}},\ \bibinfo {pages} {268} (\bibinfo {year} {2020})}\BibitemShut {NoStop}%
\bibitem [{\citenamefont {Volkoff}\ and\ \citenamefont {Martin}(2022)}]{PhysRevResearch.4.013236}%
  \BibitemOpen
  \bibfield  {author} {\bibinfo {author} {\bibfnamefont {T.~J.}\ \bibnamefont {Volkoff}}\ and\ \bibinfo {author} {\bibfnamefont {M.~J.}\ \bibnamefont {Martin}},\ }\bibfield  {title} {\bibinfo {title} {Asymptotic optimality of twist-untwist protocols for {Heisenberg} scaling in atom-based sensing},\ }\href {https://doi.org/10.1103/PhysRevResearch.4.013236} {\bibfield  {journal} {\bibinfo  {journal} {Phys. Rev. Res.}\ }\textbf {\bibinfo {volume} {4}},\ \bibinfo {pages} {013236} (\bibinfo {year} {2022})}\BibitemShut {NoStop}%
\bibitem [{\citenamefont {Li}\ \emph {et~al.}(2023{\natexlab{a}})\citenamefont {Li}, \citenamefont {R.~M.~da Silva}, \citenamefont {Kain},\ and\ \citenamefont {Shahriar}}]{PhysRevA.107.032610}%
  \BibitemOpen
  \bibfield  {author} {\bibinfo {author} {\bibfnamefont {J.}~\bibnamefont {Li}}, \bibinfo {author} {\bibfnamefont {G.}~\bibnamefont {R.~M.~da Silva}}, \bibinfo {author} {\bibfnamefont {S.}~\bibnamefont {Kain}},\ and\ \bibinfo {author} {\bibfnamefont {S.~M.}\ \bibnamefont {Shahriar}},\ }\bibfield  {title} {\bibinfo {title} {Generalized echo squeezing protocol with {near-Heisenberg-limit} sensitivity and strong robustness against detection noise and variation in squeezing parameter},\ }\href {https://doi.org/10.1103/PhysRevA.107.032610} {\bibfield  {journal} {\bibinfo  {journal} {Phys. Rev. A}\ }\textbf {\bibinfo {volume} {107}},\ \bibinfo {pages} {032610} (\bibinfo {year} {2023}{\natexlab{a}})}\BibitemShut {NoStop}%
\bibitem [{\citenamefont {Mao}\ \emph {et~al.}(2023)\citenamefont {Mao}, \citenamefont {Liu}, \citenamefont {Li}, \citenamefont {Cao}, \citenamefont {Chen}, \citenamefont {Xu}, \citenamefont {Tey}, \citenamefont {Huang},\ and\ \citenamefont {You}}]{mao2023quantum}%
  \BibitemOpen
  \bibfield  {author} {\bibinfo {author} {\bibfnamefont {T.-W.}\ \bibnamefont {Mao}}, \bibinfo {author} {\bibfnamefont {Q.}~\bibnamefont {Liu}}, \bibinfo {author} {\bibfnamefont {X.-W.}\ \bibnamefont {Li}}, \bibinfo {author} {\bibfnamefont {J.-H.}\ \bibnamefont {Cao}}, \bibinfo {author} {\bibfnamefont {F.}~\bibnamefont {Chen}}, \bibinfo {author} {\bibfnamefont {W.-X.}\ \bibnamefont {Xu}}, \bibinfo {author} {\bibfnamefont {M.~K.}\ \bibnamefont {Tey}}, \bibinfo {author} {\bibfnamefont {Y.-X.}\ \bibnamefont {Huang}},\ and\ \bibinfo {author} {\bibfnamefont {L.}~\bibnamefont {You}},\ }\bibfield  {title} {\bibinfo {title} {Quantum-enhanced sensing by echoing spin-nematic squeezing in atomic {Bose–Einstein} condensate},\ }\href {https://doi.org/10.1038/s41567-023-02168-3} {\bibfield  {journal} {\bibinfo  {journal} {Nature Physics}\ }\textbf {\bibinfo {volume} {19}},\ \bibinfo {pages} {1585} (\bibinfo {year} {2023})}\BibitemShut {NoStop}%
\bibitem [{\citenamefont {Wang}\ \emph {et~al.}(2024)\citenamefont {Wang}, \citenamefont {Filho}, \citenamefont {Agarwal},\ and\ \citenamefont {Davidovich}}]{PhysRevResearch.6.013034}%
  \BibitemOpen
  \bibfield  {author} {\bibinfo {author} {\bibfnamefont {J.}~\bibnamefont {Wang}}, \bibinfo {author} {\bibfnamefont {R.~L. d.~M.}\ \bibnamefont {Filho}}, \bibinfo {author} {\bibfnamefont {G.~S.}\ \bibnamefont {Agarwal}},\ and\ \bibinfo {author} {\bibfnamefont {L.}~\bibnamefont {Davidovich}},\ }\bibfield  {title} {\bibinfo {title} {Quantum advantage of time-reversed ancilla-based metrology of absorption parameters},\ }\href {https://doi.org/10.1103/PhysRevResearch.6.013034} {\bibfield  {journal} {\bibinfo  {journal} {Phys. Rev. Res.}\ }\textbf {\bibinfo {volume} {6}},\ \bibinfo {pages} {013034} (\bibinfo {year} {2024})}\BibitemShut {NoStop}%
\bibitem [{\citenamefont {Geier}\ \emph {et~al.}(2024)\citenamefont {Geier}, \citenamefont {Braemer}, \citenamefont {Braun}, \citenamefont {M\"ullenbach}, \citenamefont {Franz}, \citenamefont {G\"arttner}, \citenamefont {Z\"urn},\ and\ \citenamefont {Weidem\"uller}}]{PhysRevResearch.6.033197}%
  \BibitemOpen
  \bibfield  {author} {\bibinfo {author} {\bibfnamefont {S.}~\bibnamefont {Geier}}, \bibinfo {author} {\bibfnamefont {A.}~\bibnamefont {Braemer}}, \bibinfo {author} {\bibfnamefont {E.}~\bibnamefont {Braun}}, \bibinfo {author} {\bibfnamefont {M.}~\bibnamefont {M\"ullenbach}}, \bibinfo {author} {\bibfnamefont {T.}~\bibnamefont {Franz}}, \bibinfo {author} {\bibfnamefont {M.}~\bibnamefont {G\"arttner}}, \bibinfo {author} {\bibfnamefont {G.}~\bibnamefont {Z\"urn}},\ and\ \bibinfo {author} {\bibfnamefont {M.}~\bibnamefont {Weidem\"uller}},\ }\bibfield  {title} {\bibinfo {title} {Time-reversal in a dipolar quantum many-body spin system},\ }\href {https://doi.org/10.1103/PhysRevResearch.6.033197} {\bibfield  {journal} {\bibinfo  {journal} {Phys. Rev. Res.}\ }\textbf {\bibinfo {volume} {6}},\ \bibinfo {pages} {033197} (\bibinfo {year} {2024})}\BibitemShut {NoStop}%
\bibitem [{\citenamefont {Zhou}\ \emph {et~al.}(2024{\natexlab{b}})\citenamefont {Zhou}, \citenamefont {Huang},\ and\ \citenamefont {Lee}}]{PhysRevApplied.22.044066}%
  \BibitemOpen
  \bibfield  {author} {\bibinfo {author} {\bibfnamefont {J.}~\bibnamefont {Zhou}}, \bibinfo {author} {\bibfnamefont {J.}~\bibnamefont {Huang}},\ and\ \bibinfo {author} {\bibfnamefont {C.}~\bibnamefont {Lee}},\ }\bibfield  {title} {\bibinfo {title} {Scalable quantum metrology via recursive optimization},\ }\href {https://doi.org/10.1103/PhysRevApplied.22.044066} {\bibfield  {journal} {\bibinfo  {journal} {Phys. Rev. Appl.}\ }\textbf {\bibinfo {volume} {22}},\ \bibinfo {pages} {044066} (\bibinfo {year} {2024}{\natexlab{b}})}\BibitemShut {NoStop}%
\bibitem [{\citenamefont {Liu}\ \emph {et~al.}(2025)\citenamefont {Liu}, \citenamefont {Wu}, \citenamefont {Yang}, \citenamefont {Li}, \citenamefont {Zhou}, \citenamefont {Li}, \citenamefont {Chen}, \citenamefont {Yuan},\ and\ \citenamefont {Peng}}]{10.1093/nsr/nwaf091}%
  \BibitemOpen
  \bibfield  {author} {\bibinfo {author} {\bibfnamefont {R.}~\bibnamefont {Liu}}, \bibinfo {author} {\bibfnamefont {Z.}~\bibnamefont {Wu}}, \bibinfo {author} {\bibfnamefont {X.}~\bibnamefont {Yang}}, \bibinfo {author} {\bibfnamefont {Y.}~\bibnamefont {Li}}, \bibinfo {author} {\bibfnamefont {H.}~\bibnamefont {Zhou}}, \bibinfo {author} {\bibfnamefont {Z.}~\bibnamefont {Li}}, \bibinfo {author} {\bibfnamefont {Y.}~\bibnamefont {Chen}}, \bibinfo {author} {\bibfnamefont {H.}~\bibnamefont {Yuan}},\ and\ \bibinfo {author} {\bibfnamefont {X.}~\bibnamefont {Peng}},\ }\bibfield  {title} {\bibinfo {title} {Variational quantum metrology with the {Loschmidt} echo},\ }\href {https://doi.org/10.1093/nsr/nwaf091} {\bibfield  {journal} {\bibinfo  {journal} {National Science Review}\ }\textbf {\bibinfo {volume} {12}},\ \bibinfo {pages} {nwaf091} (\bibinfo {year} {2025})}\BibitemShut {NoStop}%
\bibitem [{\citenamefont {Chen}\ \emph {et~al.}(2025{\natexlab{b}})\citenamefont {Chen}, \citenamefont {Xu}, \citenamefont {Su},\ and\ \citenamefont {Li}}]{PhysRevA.111.053709}%
  \BibitemOpen
  \bibfield  {author} {\bibinfo {author} {\bibfnamefont {R.}~\bibnamefont {Chen}}, \bibinfo {author} {\bibfnamefont {P.}~\bibnamefont {Xu}}, \bibinfo {author} {\bibfnamefont {S.-L.}\ \bibnamefont {Su}},\ and\ \bibinfo {author} {\bibfnamefont {G.-X.}\ \bibnamefont {Li}},\ }\bibfield  {title} {\bibinfo {title} {Generalized nonlinear interferometer with {SU(1,1)} echo},\ }\href {https://doi.org/10.1103/PhysRevA.111.053709} {\bibfield  {journal} {\bibinfo  {journal} {Phys. Rev. A}\ }\textbf {\bibinfo {volume} {111}},\ \bibinfo {pages} {053709} (\bibinfo {year} {2025}{\natexlab{b}})}\BibitemShut {NoStop}%
\bibitem [{\citenamefont {Huang}\ \emph {et~al.}(2018{\natexlab{b}})\citenamefont {Huang}, \citenamefont {Zhuang},\ and\ \citenamefont {Lee}}]{PhysRevA.97.032116}%
  \BibitemOpen
  \bibfield  {author} {\bibinfo {author} {\bibfnamefont {J.}~\bibnamefont {Huang}}, \bibinfo {author} {\bibfnamefont {M.}~\bibnamefont {Zhuang}},\ and\ \bibinfo {author} {\bibfnamefont {C.}~\bibnamefont {Lee}},\ }\bibfield  {title} {\bibinfo {title} {{Non-Gaussian} precision metrology via driving through quantum phase transitions},\ }\href {https://doi.org/10.1103/PhysRevA.97.032116} {\bibfield  {journal} {\bibinfo  {journal} {Phys. Rev. A}\ }\textbf {\bibinfo {volume} {97}},\ \bibinfo {pages} {032116} (\bibinfo {year} {2018}{\natexlab{b}})}\BibitemShut {NoStop}%
\bibitem [{\citenamefont {Zhuang}\ \emph {et~al.}(2021)\citenamefont {Zhuang}, \citenamefont {Huo}, \citenamefont {Qiu}, \citenamefont {Liu}, \citenamefont {Huang},\ and\ \citenamefont {Lee}}]{PhysRevApplied.16.064056}%
  \BibitemOpen
  \bibfield  {author} {\bibinfo {author} {\bibfnamefont {M.}~\bibnamefont {Zhuang}}, \bibinfo {author} {\bibfnamefont {H.}~\bibnamefont {Huo}}, \bibinfo {author} {\bibfnamefont {Y.}~\bibnamefont {Qiu}}, \bibinfo {author} {\bibfnamefont {W.}~\bibnamefont {Liu}}, \bibinfo {author} {\bibfnamefont {J.}~\bibnamefont {Huang}},\ and\ \bibinfo {author} {\bibfnamefont {C.}~\bibnamefont {Lee}},\ }\bibfield  {title} {\bibinfo {title} {Heisenberg-limited frequency estimation via driving through quantum phase transitions},\ }\href {https://doi.org/10.1103/PhysRevApplied.16.064056} {\bibfield  {journal} {\bibinfo  {journal} {Phys. Rev. Appl.}\ }\textbf {\bibinfo {volume} {16}},\ \bibinfo {pages} {064056} (\bibinfo {year} {2021})}\BibitemShut {NoStop}%
\bibitem [{\citenamefont {Liu}\ \emph {et~al.}(2022)\citenamefont {Liu}, \citenamefont {Wu}, \citenamefont {Cao}, \citenamefont {Mao}, \citenamefont {Li}, \citenamefont {Guo}, \citenamefont {Tey},\ and\ \citenamefont {You}}]{liu2022nonlinear}%
  \BibitemOpen
  \bibfield  {author} {\bibinfo {author} {\bibfnamefont {Q.}~\bibnamefont {Liu}}, \bibinfo {author} {\bibfnamefont {L.-N.}\ \bibnamefont {Wu}}, \bibinfo {author} {\bibfnamefont {J.-H.}\ \bibnamefont {Cao}}, \bibinfo {author} {\bibfnamefont {T.-W.}\ \bibnamefont {Mao}}, \bibinfo {author} {\bibfnamefont {X.-W.}\ \bibnamefont {Li}}, \bibinfo {author} {\bibfnamefont {S.-F.}\ \bibnamefont {Guo}}, \bibinfo {author} {\bibfnamefont {M.~K.}\ \bibnamefont {Tey}},\ and\ \bibinfo {author} {\bibfnamefont {L.}~\bibnamefont {You}},\ }\bibfield  {title} {\bibinfo {title} {Nonlinear interferometry beyond classical limit enabled by cyclic dynamics},\ }\href {https://doi.org/10.1038/s41567-021-01441-7} {\bibfield  {journal} {\bibinfo  {journal} {Nature Physics}\ }\textbf {\bibinfo {volume} {18}},\ \bibinfo {pages} {167} (\bibinfo {year} {2022})}\BibitemShut {NoStop}%
\bibitem [{\citenamefont {Liu}\ \emph {et~al.}(2023)\citenamefont {Liu}, \citenamefont {Mao}, \citenamefont {Xue}, \citenamefont {Wu},\ and\ \citenamefont {You}}]{PhysRevA.107.052613}%
  \BibitemOpen
  \bibfield  {author} {\bibinfo {author} {\bibfnamefont {Q.}~\bibnamefont {Liu}}, \bibinfo {author} {\bibfnamefont {T.-W.}\ \bibnamefont {Mao}}, \bibinfo {author} {\bibfnamefont {M.}~\bibnamefont {Xue}}, \bibinfo {author} {\bibfnamefont {L.-N.}\ \bibnamefont {Wu}},\ and\ \bibinfo {author} {\bibfnamefont {L.}~\bibnamefont {You}},\ }\bibfield  {title} {\bibinfo {title} {Cyclic nonlinear interferometry with entangled {non-Gaussian} spin states},\ }\href {https://doi.org/10.1103/PhysRevA.107.052613} {\bibfield  {journal} {\bibinfo  {journal} {Phys. Rev. A}\ }\textbf {\bibinfo {volume} {107}},\ \bibinfo {pages} {052613} (\bibinfo {year} {2023})}\BibitemShut {NoStop}%
\bibitem [{\citenamefont {Huang}\ \emph {et~al.}(2021)\citenamefont {Huang}, \citenamefont {Chen}, \citenamefont {Li}, \citenamefont {Li}, \citenamefont {L{\"u}},\ and\ \citenamefont {Liu}}]{huang2021dynamic}%
  \BibitemOpen
  \bibfield  {author} {\bibinfo {author} {\bibfnamefont {L.-G.}\ \bibnamefont {Huang}}, \bibinfo {author} {\bibfnamefont {F.}~\bibnamefont {Chen}}, \bibinfo {author} {\bibfnamefont {X.}~\bibnamefont {Li}}, \bibinfo {author} {\bibfnamefont {Y.}~\bibnamefont {Li}}, \bibinfo {author} {\bibfnamefont {R.}~\bibnamefont {L{\"u}}},\ and\ \bibinfo {author} {\bibfnamefont {Y.-C.}\ \bibnamefont {Liu}},\ }\bibfield  {title} {\bibinfo {title} {Dynamic synthesis of {Heisenberg}-limited spin squeezing},\ }\href {https://doi.org/10.1038/s41534-021-00505-z} {\bibfield  {journal} {\bibinfo  {journal} {npj Quantum Information}\ }\textbf {\bibinfo {volume} {7}},\ \bibinfo {pages} {168} (\bibinfo {year} {2021})}\BibitemShut {NoStop}%
\bibitem [{\citenamefont {Hu}\ \emph {et~al.}(2024)\citenamefont {Hu}, \citenamefont {Li}, \citenamefont {Zhang}, \citenamefont {Zhang}, \citenamefont {Huang},\ and\ \citenamefont {Liu}}]{hu2024nonlinear}%
  \BibitemOpen
  \bibfield  {author} {\bibinfo {author} {\bibfnamefont {Z.}~\bibnamefont {Hu}}, \bibinfo {author} {\bibfnamefont {Q.}~\bibnamefont {Li}}, \bibinfo {author} {\bibfnamefont {X.}~\bibnamefont {Zhang}}, \bibinfo {author} {\bibfnamefont {H.-B.}\ \bibnamefont {Zhang}}, \bibinfo {author} {\bibfnamefont {L.-G.}\ \bibnamefont {Huang}},\ and\ \bibinfo {author} {\bibfnamefont {Y.-C.}\ \bibnamefont {Liu}},\ }\bibfield  {title} {\bibinfo {title} {Nonlinear time-reversal interferometry with arbitrary quadratic collective-spin interaction},\ }\href {https://cpb.iphy.ac.cn/CN/abstract/article_126964.shtml} {\bibfield  {journal} {\bibinfo  {journal} {Chinese Physics B}\ }\textbf {\bibinfo {volume} {33}},\ \bibinfo {pages} {080601} (\bibinfo {year} {2024})}\BibitemShut {NoStop}%
\bibitem [{\citenamefont {Ma}\ \emph {et~al.}(2024)\citenamefont {Ma}, \citenamefont {Zhou}, \citenamefont {Huang},\ and\ \citenamefont {Lee}}]{PhysRevA.110.022407}%
  \BibitemOpen
  \bibfield  {author} {\bibinfo {author} {\bibfnamefont {J.}~\bibnamefont {Ma}}, \bibinfo {author} {\bibfnamefont {J.}~\bibnamefont {Zhou}}, \bibinfo {author} {\bibfnamefont {J.}~\bibnamefont {Huang}},\ and\ \bibinfo {author} {\bibfnamefont {C.}~\bibnamefont {Lee}},\ }\bibfield  {title} {\bibinfo {title} {Phase amplification via synthetic two-axis-twisting echo from interaction-fixed one-axis twisting},\ }\href {https://doi.org/10.1103/PhysRevA.110.022407} {\bibfield  {journal} {\bibinfo  {journal} {Phys. Rev. A}\ }\textbf {\bibinfo {volume} {110}},\ \bibinfo {pages} {022407} (\bibinfo {year} {2024})}\BibitemShut {NoStop}%
\bibitem [{\citenamefont {Gao}\ \emph {et~al.}(2025)\citenamefont {Gao}, \citenamefont {Martin}, \citenamefont {Hughes}, \citenamefont {Leitao}, \citenamefont {Put}, \citenamefont {Zhou}, \citenamefont {Koyluoglu}, \citenamefont {Meynell}, \citenamefont {Jayich}, \citenamefont {Park} \emph {et~al.}}]{gao2025signal}%
  \BibitemOpen
  \bibfield  {author} {\bibinfo {author} {\bibfnamefont {H.}~\bibnamefont {Gao}}, \bibinfo {author} {\bibfnamefont {L.~S.}\ \bibnamefont {Martin}}, \bibinfo {author} {\bibfnamefont {L.~B.}\ \bibnamefont {Hughes}}, \bibinfo {author} {\bibfnamefont {N.~T.}\ \bibnamefont {Leitao}}, \bibinfo {author} {\bibfnamefont {P.}~\bibnamefont {Put}}, \bibinfo {author} {\bibfnamefont {H.}~\bibnamefont {Zhou}}, \bibinfo {author} {\bibfnamefont {N.~U.}\ \bibnamefont {Koyluoglu}}, \bibinfo {author} {\bibfnamefont {S.~A.}\ \bibnamefont {Meynell}}, \bibinfo {author} {\bibfnamefont {A.~C.~B.}\ \bibnamefont {Jayich}}, \bibinfo {author} {\bibfnamefont {H.}~\bibnamefont {Park}}, \emph {et~al.},\ }\bibfield  {title} {\bibinfo {title} {Signal amplification in a solid-state sensor through asymmetric many-body echo},\ }\href {https://doi.org/10.1038/s41586-025-09452-7} {\bibfield  {journal} {\bibinfo  {journal} {Nature}\ }\textbf {\bibinfo {volume} {646}},\ \bibinfo {pages} {68} (\bibinfo {year} {2025})}\BibitemShut {NoStop}%
\bibitem [{\citenamefont {Ma}\ \emph {et~al.}(2025)\citenamefont {Ma}, \citenamefont {Shen}, \citenamefont {Huang},\ and\ \citenamefont {Lee}}]{32n5-mhk1}%
  \BibitemOpen
  \bibfield  {author} {\bibinfo {author} {\bibfnamefont {J.}~\bibnamefont {Ma}}, \bibinfo {author} {\bibfnamefont {Y.}~\bibnamefont {Shen}}, \bibinfo {author} {\bibfnamefont {J.}~\bibnamefont {Huang}},\ and\ \bibinfo {author} {\bibfnamefont {C.}~\bibnamefont {Lee}},\ }\bibfield  {title} {\bibinfo {title} {Quantum metrology via {Floquet}-engineered two-axis twisting and turning dynamics},\ }\href {https://doi.org/10.1103/32n5-mhk1} {\bibfield  {journal} {\bibinfo  {journal} {Phys. Rev. A}\ }\textbf {\bibinfo {volume} {112}},\ \bibinfo {pages} {L040602} (\bibinfo {year} {2025})}\BibitemShut {NoStop}%
\bibitem [{\citenamefont {Colombo}\ \emph {et~al.}(2022)\citenamefont {Colombo}, \citenamefont {Pedrozo-Peñafiel}, \citenamefont {Adiyatullin}, \citenamefont {Li}, \citenamefont {Mendez}, \citenamefont {Shu},\ and\ \citenamefont {Vuletić}}]{colombo_time-reversal-based_2022}%
  \BibitemOpen
  \bibfield  {author} {\bibinfo {author} {\bibfnamefont {S.}~\bibnamefont {Colombo}}, \bibinfo {author} {\bibfnamefont {E.}~\bibnamefont {Pedrozo-Peñafiel}}, \bibinfo {author} {\bibfnamefont {A.~F.}\ \bibnamefont {Adiyatullin}}, \bibinfo {author} {\bibfnamefont {Z.}~\bibnamefont {Li}}, \bibinfo {author} {\bibfnamefont {E.}~\bibnamefont {Mendez}}, \bibinfo {author} {\bibfnamefont {C.}~\bibnamefont {Shu}},\ and\ \bibinfo {author} {\bibfnamefont {V.}~\bibnamefont {Vuletić}},\ }\bibfield  {title} {\bibinfo {title} {Time-reversal-based quantum metrology with many-body entangled states},\ }\href {https://doi.org/10.1038/s41567-022-01653-5} {\bibfield  {journal} {\bibinfo  {journal} {Nature Physics}\ }\textbf {\bibinfo {volume} {18}},\ \bibinfo {pages} {925} (\bibinfo {year} {2022})}\BibitemShut {NoStop}%
\bibitem [{\citenamefont {Li}\ \emph {et~al.}(2023{\natexlab{b}})\citenamefont {Li}, \citenamefont {null}, \citenamefont {Colombo}, \citenamefont {Shu}, \citenamefont {Velez}, \citenamefont {Pilatowsky-Cameo}, \citenamefont {Schmied}, \citenamefont {Choi}, \citenamefont {Lukin}, \citenamefont {Pedrozo-Peñafiel},\ and\ \citenamefont {Vuletić}}]{li2023improving}%
  \BibitemOpen
  \bibfield  {author} {\bibinfo {author} {\bibfnamefont {Z.}~\bibnamefont {Li}}, \bibinfo {author} {\bibnamefont {null}}, \bibinfo {author} {\bibfnamefont {S.}~\bibnamefont {Colombo}}, \bibinfo {author} {\bibfnamefont {C.}~\bibnamefont {Shu}}, \bibinfo {author} {\bibfnamefont {G.}~\bibnamefont {Velez}}, \bibinfo {author} {\bibfnamefont {S.}~\bibnamefont {Pilatowsky-Cameo}}, \bibinfo {author} {\bibfnamefont {R.}~\bibnamefont {Schmied}}, \bibinfo {author} {\bibfnamefont {S.}~\bibnamefont {Choi}}, \bibinfo {author} {\bibfnamefont {M.}~\bibnamefont {Lukin}}, \bibinfo {author} {\bibfnamefont {E.}~\bibnamefont {Pedrozo-Peñafiel}},\ and\ \bibinfo {author} {\bibfnamefont {V.}~\bibnamefont {Vuletić}},\ }\bibfield  {title} {\bibinfo {title} {Improving metrology with quantum scrambling},\ }\href {https://doi.org/10.1126/science.adg9500} {\bibfield  {journal} {\bibinfo  {journal} {Science}\ }\textbf {\bibinfo {volume} {380}},\ \bibinfo {pages} {1381} (\bibinfo {year} {2023}{\natexlab{b}})}\BibitemShut {NoStop}%
\bibitem [{\citenamefont {Zaporski}\ \emph {et~al.}(2025)\citenamefont {Zaporski}, \citenamefont {Liu}, \citenamefont {Velez}, \citenamefont {Radzihovsky}, \citenamefont {Li}, \citenamefont {Colombo}, \citenamefont {Pedrozo-Pe{\~n}afiel},\ and\ \citenamefont {Vuleti{\'c}}}]{zaporski2025quantum}%
  \BibitemOpen
  \bibfield  {author} {\bibinfo {author} {\bibfnamefont {L.}~\bibnamefont {Zaporski}}, \bibinfo {author} {\bibfnamefont {Q.}~\bibnamefont {Liu}}, \bibinfo {author} {\bibfnamefont {G.}~\bibnamefont {Velez}}, \bibinfo {author} {\bibfnamefont {M.}~\bibnamefont {Radzihovsky}}, \bibinfo {author} {\bibfnamefont {Z.}~\bibnamefont {Li}}, \bibinfo {author} {\bibfnamefont {S.}~\bibnamefont {Colombo}}, \bibinfo {author} {\bibfnamefont {E.}~\bibnamefont {Pedrozo-Pe{\~n}afiel}},\ and\ \bibinfo {author} {\bibfnamefont {V.}~\bibnamefont {Vuleti{\'c}}},\ }\bibfield  {title} {\bibinfo {title} {Quantum-amplified global-phase spectroscopy on an optical clock transition},\ }\href {https://doi.org/10.1038/s41586-025-09578-8} {\bibfield  {journal} {\bibinfo  {journal} {Nature}\ }\textbf {\bibinfo {volume} {646}},\ \bibinfo {pages} {309} (\bibinfo {year} {2025})}\BibitemShut {NoStop}%
\bibitem [{\citenamefont {Linnemann}\ \emph {et~al.}(2016)\citenamefont {Linnemann}, \citenamefont {Strobel}, \citenamefont {Muessel}, \citenamefont {Schulz}, \citenamefont {Lewis-Swan}, \citenamefont {Kheruntsyan},\ and\ \citenamefont {Oberthaler}}]{PhysRevLett.117.013001}%
  \BibitemOpen
  \bibfield  {author} {\bibinfo {author} {\bibfnamefont {D.}~\bibnamefont {Linnemann}}, \bibinfo {author} {\bibfnamefont {H.}~\bibnamefont {Strobel}}, \bibinfo {author} {\bibfnamefont {W.}~\bibnamefont {Muessel}}, \bibinfo {author} {\bibfnamefont {J.}~\bibnamefont {Schulz}}, \bibinfo {author} {\bibfnamefont {R.~J.}\ \bibnamefont {Lewis-Swan}}, \bibinfo {author} {\bibfnamefont {K.~V.}\ \bibnamefont {Kheruntsyan}},\ and\ \bibinfo {author} {\bibfnamefont {M.~K.}\ \bibnamefont {Oberthaler}},\ }\bibfield  {title} {\bibinfo {title} {Quantum-enhanced sensing based on time reversal of nonlinear dynamics},\ }\href {https://doi.org/10.1103/PhysRevLett.117.013001} {\bibfield  {journal} {\bibinfo  {journal} {Phys. Rev. Lett.}\ }\textbf {\bibinfo {volume} {117}},\ \bibinfo {pages} {013001} (\bibinfo {year} {2016})}\BibitemShut {NoStop}%
\bibitem [{\citenamefont {Linnemann}\ \emph {et~al.}(2017)\citenamefont {Linnemann}, \citenamefont {Schulz}, \citenamefont {Muessel}, \citenamefont {Kunkel}, \citenamefont {Prüfer}, \citenamefont {Frölian}, \citenamefont {Strobel},\ and\ \citenamefont {Oberthaler}}]{Linnemann_2017}%
  \BibitemOpen
  \bibfield  {author} {\bibinfo {author} {\bibfnamefont {D.}~\bibnamefont {Linnemann}}, \bibinfo {author} {\bibfnamefont {J.}~\bibnamefont {Schulz}}, \bibinfo {author} {\bibfnamefont {W.}~\bibnamefont {Muessel}}, \bibinfo {author} {\bibfnamefont {P.}~\bibnamefont {Kunkel}}, \bibinfo {author} {\bibfnamefont {M.}~\bibnamefont {Prüfer}}, \bibinfo {author} {\bibfnamefont {A.}~\bibnamefont {Frölian}}, \bibinfo {author} {\bibfnamefont {H.}~\bibnamefont {Strobel}},\ and\ \bibinfo {author} {\bibfnamefont {M.~K.}\ \bibnamefont {Oberthaler}},\ }\bibfield  {title} {\bibinfo {title} {Active {SU(1,1)} atom interferometry},\ }\href {https://doi.org/10.1088/2058-9565/aa802c} {\bibfield  {journal} {\bibinfo  {journal} {Quantum Science and Technology}\ }\textbf {\bibinfo {volume} {2}},\ \bibinfo {pages} {044009} (\bibinfo {year} {2017})}\BibitemShut {NoStop}%
\bibitem [{\citenamefont {Hudelist}\ \emph {et~al.}(2014)\citenamefont {Hudelist}, \citenamefont {Kong}, \citenamefont {Liu}, \citenamefont {Jing}, \citenamefont {Ou},\ and\ \citenamefont {Zhang}}]{hudelist2014quantum}%
  \BibitemOpen
  \bibfield  {author} {\bibinfo {author} {\bibfnamefont {F.}~\bibnamefont {Hudelist}}, \bibinfo {author} {\bibfnamefont {J.}~\bibnamefont {Kong}}, \bibinfo {author} {\bibfnamefont {C.}~\bibnamefont {Liu}}, \bibinfo {author} {\bibfnamefont {J.}~\bibnamefont {Jing}}, \bibinfo {author} {\bibfnamefont {Z.}~\bibnamefont {Ou}},\ and\ \bibinfo {author} {\bibfnamefont {W.}~\bibnamefont {Zhang}},\ }\bibfield  {title} {\bibinfo {title} {Quantum metrology with parametric amplifier-based photon correlation interferometers},\ }\href {https://doi.org/10.1038/ncomms4049} {\bibfield  {journal} {\bibinfo  {journal} {Nature communications}\ }\textbf {\bibinfo {volume} {5}},\ \bibinfo {pages} {3049} (\bibinfo {year} {2014})}\BibitemShut {NoStop}%
\bibitem [{\citenamefont {Manceau}\ \emph {et~al.}(2017)\citenamefont {Manceau}, \citenamefont {Leuchs}, \citenamefont {Khalili},\ and\ \citenamefont {Chekhova}}]{PhysRevLett.119.223604}%
  \BibitemOpen
  \bibfield  {author} {\bibinfo {author} {\bibfnamefont {M.}~\bibnamefont {Manceau}}, \bibinfo {author} {\bibfnamefont {G.}~\bibnamefont {Leuchs}}, \bibinfo {author} {\bibfnamefont {F.}~\bibnamefont {Khalili}},\ and\ \bibinfo {author} {\bibfnamefont {M.}~\bibnamefont {Chekhova}},\ }\bibfield  {title} {\bibinfo {title} {Detection loss tolerant supersensitive phase measurement with an {SU(1,1)} interferometer},\ }\href {https://doi.org/10.1103/PhysRevLett.119.223604} {\bibfield  {journal} {\bibinfo  {journal} {Phys. Rev. Lett.}\ }\textbf {\bibinfo {volume} {119}},\ \bibinfo {pages} {223604} (\bibinfo {year} {2017})}\BibitemShut {NoStop}%
\bibitem [{\citenamefont {Tian}\ \emph {et~al.}(2023)\citenamefont {Tian}, \citenamefont {Yao}, \citenamefont {Wu}, \citenamefont {Wang}, \citenamefont {Shen}, \citenamefont {Zheng},\ and\ \citenamefont {Peng}}]{Tian:23}%
  \BibitemOpen
  \bibfield  {author} {\bibinfo {author} {\bibfnamefont {L.}~\bibnamefont {Tian}}, \bibinfo {author} {\bibfnamefont {W.}~\bibnamefont {Yao}}, \bibinfo {author} {\bibfnamefont {Y.}~\bibnamefont {Wu}}, \bibinfo {author} {\bibfnamefont {Q.}~\bibnamefont {Wang}}, \bibinfo {author} {\bibfnamefont {H.}~\bibnamefont {Shen}}, \bibinfo {author} {\bibfnamefont {Y.}~\bibnamefont {Zheng}},\ and\ \bibinfo {author} {\bibfnamefont {K.}~\bibnamefont {Peng}},\ }\bibfield  {title} {\bibinfo {title} {Loss-tolerant and quantum-enhanced interferometer by reversed squeezing processes},\ }\href {https://doi.org/10.1364/OL.487355} {\bibfield  {journal} {\bibinfo  {journal} {Opt. Lett.}\ }\textbf {\bibinfo {volume} {48}},\ \bibinfo {pages} {3909} (\bibinfo {year} {2023})}\BibitemShut {NoStop}%
\bibitem [{\citenamefont {G{\"a}rttner}\ \emph {et~al.}(2017)\citenamefont {G{\"a}rttner}, \citenamefont {Bohnet}, \citenamefont {Safavi-Naini}, \citenamefont {Wall}, \citenamefont {Bollinger},\ and\ \citenamefont {Rey}}]{garttner2017measuring}%
  \BibitemOpen
  \bibfield  {author} {\bibinfo {author} {\bibfnamefont {M.}~\bibnamefont {G{\"a}rttner}}, \bibinfo {author} {\bibfnamefont {J.~G.}\ \bibnamefont {Bohnet}}, \bibinfo {author} {\bibfnamefont {A.}~\bibnamefont {Safavi-Naini}}, \bibinfo {author} {\bibfnamefont {M.~L.}\ \bibnamefont {Wall}}, \bibinfo {author} {\bibfnamefont {J.~J.}\ \bibnamefont {Bollinger}},\ and\ \bibinfo {author} {\bibfnamefont {A.~M.}\ \bibnamefont {Rey}},\ }\bibfield  {title} {\bibinfo {title} {Measuring out-of-time-order correlations and multiple quantum spectra in a trapped-ion quantum magnet},\ }\href {https://doi.org/10.1038/nphys4119} {\bibfield  {journal} {\bibinfo  {journal} {Nature Physics}\ }\textbf {\bibinfo {volume} {13}},\ \bibinfo {pages} {781} (\bibinfo {year} {2017})}\BibitemShut {NoStop}%
\bibitem [{\citenamefont {Burd}\ \emph {et~al.}(2019)\citenamefont {Burd}, \citenamefont {Srinivas}, \citenamefont {Bollinger}, \citenamefont {Wilson}, \citenamefont {Wineland}, \citenamefont {Leibfried}, \citenamefont {Slichter},\ and\ \citenamefont {Allcock}}]{doi:10.1126/science.aaw2884}%
  \BibitemOpen
  \bibfield  {author} {\bibinfo {author} {\bibfnamefont {S.~C.}\ \bibnamefont {Burd}}, \bibinfo {author} {\bibfnamefont {R.}~\bibnamefont {Srinivas}}, \bibinfo {author} {\bibfnamefont {J.~J.}\ \bibnamefont {Bollinger}}, \bibinfo {author} {\bibfnamefont {A.~C.}\ \bibnamefont {Wilson}}, \bibinfo {author} {\bibfnamefont {D.~J.}\ \bibnamefont {Wineland}}, \bibinfo {author} {\bibfnamefont {D.}~\bibnamefont {Leibfried}}, \bibinfo {author} {\bibfnamefont {D.~H.}\ \bibnamefont {Slichter}},\ and\ \bibinfo {author} {\bibfnamefont {D.~T.~C.}\ \bibnamefont {Allcock}},\ }\bibfield  {title} {\bibinfo {title} {Quantum amplification of mechanical oscillator motion},\ }\href {https://doi.org/10.1126/science.aaw2884} {\bibfield  {journal} {\bibinfo  {journal} {Science}\ }\textbf {\bibinfo {volume} {364}},\ \bibinfo {pages} {1163} (\bibinfo {year} {2019})}\BibitemShut {NoStop}%
\bibitem [{\citenamefont {Gilmore}\ \emph {et~al.}(2021)\citenamefont {Gilmore}, \citenamefont {Affolter}, \citenamefont {Lewis-Swan}, \citenamefont {Barberena}, \citenamefont {Jordan}, \citenamefont {Rey},\ and\ \citenamefont {Bollinger}}]{doi:10.1126/science.abi5226}%
  \BibitemOpen
  \bibfield  {author} {\bibinfo {author} {\bibfnamefont {K.~A.}\ \bibnamefont {Gilmore}}, \bibinfo {author} {\bibfnamefont {M.}~\bibnamefont {Affolter}}, \bibinfo {author} {\bibfnamefont {R.~J.}\ \bibnamefont {Lewis-Swan}}, \bibinfo {author} {\bibfnamefont {D.}~\bibnamefont {Barberena}}, \bibinfo {author} {\bibfnamefont {E.}~\bibnamefont {Jordan}}, \bibinfo {author} {\bibfnamefont {A.~M.}\ \bibnamefont {Rey}},\ and\ \bibinfo {author} {\bibfnamefont {J.~J.}\ \bibnamefont {Bollinger}},\ }\bibfield  {title} {\bibinfo {title} {Quantum-enhanced sensing of displacements and electric fields with two-dimensional trapped-ion crystals},\ }\href {https://doi.org/10.1126/science.abi5226} {\bibfield  {journal} {\bibinfo  {journal} {Science}\ }\textbf {\bibinfo {volume} {373}},\ \bibinfo {pages} {673} (\bibinfo {year} {2021})}\BibitemShut {NoStop}%
\bibitem [{\citenamefont {Raghavan}\ \emph {et~al.}(1999)\citenamefont {Raghavan}, \citenamefont {Smerzi}, \citenamefont {Fantoni},\ and\ \citenamefont {Shenoy}}]{PhysRevA.59.620}%
  \BibitemOpen
  \bibfield  {author} {\bibinfo {author} {\bibfnamefont {S.}~\bibnamefont {Raghavan}}, \bibinfo {author} {\bibfnamefont {A.}~\bibnamefont {Smerzi}}, \bibinfo {author} {\bibfnamefont {S.}~\bibnamefont {Fantoni}},\ and\ \bibinfo {author} {\bibfnamefont {S.~R.}\ \bibnamefont {Shenoy}},\ }\bibfield  {title} {\bibinfo {title} {Coherent oscillations between two weakly coupled {Bose-Einstein} condensates: {Josephson} effects, $\ensuremath{\pi}$ oscillations, and macroscopic quantum self-trapping},\ }\href {https://doi.org/10.1103/PhysRevA.59.620} {\bibfield  {journal} {\bibinfo  {journal} {Phys. Rev. A}\ }\textbf {\bibinfo {volume} {59}},\ \bibinfo {pages} {620} (\bibinfo {year} {1999})}\BibitemShut {NoStop}%
\bibitem [{\citenamefont {Lee}(2006)}]{PhysRevLett.97.150402}%
  \BibitemOpen
  \bibfield  {author} {\bibinfo {author} {\bibfnamefont {C.}~\bibnamefont {Lee}},\ }\bibfield  {title} {\bibinfo {title} {Adiabatic {Mach-Zehnder} interferometry on a quantized {Bose-Josephson} junction},\ }\href {https://doi.org/10.1103/PhysRevLett.97.150402} {\bibfield  {journal} {\bibinfo  {journal} {Phys. Rev. Lett.}\ }\textbf {\bibinfo {volume} {97}},\ \bibinfo {pages} {150402} (\bibinfo {year} {2006})}\BibitemShut {NoStop}%
\bibitem [{\citenamefont {Lee}\ \emph {et~al.}(2012)\citenamefont {Lee}, \citenamefont {Huang}, \citenamefont {Deng}, \citenamefont {Dai},\ and\ \citenamefont {Xu}}]{lee2012nonlinear}%
  \BibitemOpen
  \bibfield  {author} {\bibinfo {author} {\bibfnamefont {C.}~\bibnamefont {Lee}}, \bibinfo {author} {\bibfnamefont {J.}~\bibnamefont {Huang}}, \bibinfo {author} {\bibfnamefont {H.}~\bibnamefont {Deng}}, \bibinfo {author} {\bibfnamefont {H.}~\bibnamefont {Dai}},\ and\ \bibinfo {author} {\bibfnamefont {J.}~\bibnamefont {Xu}},\ }\bibfield  {title} {\bibinfo {title} {Nonlinear quantum interferometry with {Bose} condensed atoms},\ }\href {https://doi.org/10.1007/s11467-011-0228-6} {\bibfield  {journal} {\bibinfo  {journal} {Frontiers of physics}\ }\textbf {\bibinfo {volume} {7}},\ \bibinfo {pages} {109} (\bibinfo {year} {2012})}\BibitemShut {NoStop}%
\bibitem [{\citenamefont {Liu}\ \emph {et~al.}(2021)\citenamefont {Liu}, \citenamefont {Zhuang}, \citenamefont {Zhu}, \citenamefont {Huang},\ and\ \citenamefont {Lee}}]{PhysRevA.103.023309}%
  \BibitemOpen
  \bibfield  {author} {\bibinfo {author} {\bibfnamefont {W.}~\bibnamefont {Liu}}, \bibinfo {author} {\bibfnamefont {M.}~\bibnamefont {Zhuang}}, \bibinfo {author} {\bibfnamefont {B.}~\bibnamefont {Zhu}}, \bibinfo {author} {\bibfnamefont {J.}~\bibnamefont {Huang}},\ and\ \bibinfo {author} {\bibfnamefont {C.}~\bibnamefont {Lee}},\ }\bibfield  {title} {\bibinfo {title} {Quantum metrology via chaos in a driven {Bose-Josephson} system},\ }\href {https://doi.org/10.1103/PhysRevA.103.023309} {\bibfield  {journal} {\bibinfo  {journal} {Phys. Rev. A}\ }\textbf {\bibinfo {volume} {103}},\ \bibinfo {pages} {023309} (\bibinfo {year} {2021})}\BibitemShut {NoStop}%
\bibitem [{\citenamefont {Yurke}\ \emph {et~al.}(1986)\citenamefont {Yurke}, \citenamefont {McCall},\ and\ \citenamefont {Klauder}}]{PhysRevA.33.4033}%
  \BibitemOpen
  \bibfield  {author} {\bibinfo {author} {\bibfnamefont {B.}~\bibnamefont {Yurke}}, \bibinfo {author} {\bibfnamefont {S.~L.}\ \bibnamefont {McCall}},\ and\ \bibinfo {author} {\bibfnamefont {J.~R.}\ \bibnamefont {Klauder}},\ }\bibfield  {title} {\bibinfo {title} {{SU(2)} and {SU(1,1)} interferometers},\ }\href {https://doi.org/10.1103/PhysRevA.33.4033} {\bibfield  {journal} {\bibinfo  {journal} {Phys. Rev. A}\ }\textbf {\bibinfo {volume} {33}},\ \bibinfo {pages} {4033} (\bibinfo {year} {1986})}\BibitemShut {NoStop}%
\bibitem [{\citenamefont {Pezz\`e}\ and\ \citenamefont {Smerzi}(2020)}]{PhysRevLett.125.210503}%
  \BibitemOpen
  \bibfield  {author} {\bibinfo {author} {\bibfnamefont {L.}~\bibnamefont {Pezz\`e}}\ and\ \bibinfo {author} {\bibfnamefont {A.}~\bibnamefont {Smerzi}},\ }\bibfield  {title} {\bibinfo {title} {Heisenberg-limited noisy atomic clock using a hybrid coherent and squeezed state protocol},\ }\href {https://doi.org/10.1103/PhysRevLett.125.210503} {\bibfield  {journal} {\bibinfo  {journal} {Phys. Rev. Lett.}\ }\textbf {\bibinfo {volume} {125}},\ \bibinfo {pages} {210503} (\bibinfo {year} {2020})}\BibitemShut {NoStop}%
\bibitem [{SM()}]{SM}%
  \BibitemOpen
  \href@noop {} {}\bibinfo {note} {See Supplemental Material for details on: (i) Derivation of Floquet Hamiltonian; (ii) Impact of nonlinear interaction on signal accumulation; (iii) Entangled state preparation and interaction-based readout via AIQM; (iv) Detection noise and (v) Experimental feasibility.}\BibitemShut {Stop}%
\bibitem [{\citenamefont {Van-Brunt}\ and\ \citenamefont {Visser}(2015)}]{Van-Brunt_2015}%
  \BibitemOpen
  \bibfield  {author} {\bibinfo {author} {\bibfnamefont {A.}~\bibnamefont {Van-Brunt}}\ and\ \bibinfo {author} {\bibfnamefont {M.}~\bibnamefont {Visser}},\ }\bibfield  {title} {\bibinfo {title} {Special-case closed form of the {Baker–Campbell–Hausdorff} formula},\ }\href {https://doi.org/10.1088/1751-8113/48/22/225207} {\bibfield  {journal} {\bibinfo  {journal} {Journal of Physics A: Mathematical and Theoretical}\ }\textbf {\bibinfo {volume} {48}},\ \bibinfo {pages} {225207} (\bibinfo {year} {2015})}\BibitemShut {NoStop}%
\bibitem [{\citenamefont {Zhang}\ \emph {et~al.}(2014)\citenamefont {Zhang}, \citenamefont {Zhou}, \citenamefont {Guo},\ and\ \citenamefont {Zhou}}]{PhysRevA.90.013604}%
  \BibitemOpen
  \bibfield  {author} {\bibinfo {author} {\bibfnamefont {J.-Y.}\ \bibnamefont {Zhang}}, \bibinfo {author} {\bibfnamefont {X.-F.}\ \bibnamefont {Zhou}}, \bibinfo {author} {\bibfnamefont {G.-C.}\ \bibnamefont {Guo}},\ and\ \bibinfo {author} {\bibfnamefont {Z.-W.}\ \bibnamefont {Zhou}},\ }\bibfield  {title} {\bibinfo {title} {Dynamical spin squeezing via a higher-order {Trotter-Suzuki} approximation},\ }\href {https://doi.org/10.1103/PhysRevA.90.013604} {\bibfield  {journal} {\bibinfo  {journal} {Phys. Rev. A}\ }\textbf {\bibinfo {volume} {90}},\ \bibinfo {pages} {013604} (\bibinfo {year} {2014})}\BibitemShut {NoStop}%
\bibitem [{\citenamefont {Ma}\ \emph {et~al.}(2011)\citenamefont {Ma}, \citenamefont {Wang}, \citenamefont {Sun},\ and\ \citenamefont {Nori}}]{MA201189}%
  \BibitemOpen
  \bibfield  {author} {\bibinfo {author} {\bibfnamefont {J.}~\bibnamefont {Ma}}, \bibinfo {author} {\bibfnamefont {X.}~\bibnamefont {Wang}}, \bibinfo {author} {\bibfnamefont {C.}~\bibnamefont {Sun}},\ and\ \bibinfo {author} {\bibfnamefont {F.}~\bibnamefont {Nori}},\ }\bibfield  {title} {\bibinfo {title} {Quantum spin squeezing},\ }\href {https://doi.org/https://doi.org/10.1016/j.physrep.2011.08.003} {\bibfield  {journal} {\bibinfo  {journal} {Physics Reports}\ }\textbf {\bibinfo {volume} {509}},\ \bibinfo {pages} {89} (\bibinfo {year} {2011})}\BibitemShut {NoStop}%
\bibitem [{\citenamefont {Huang}\ \emph {et~al.}(2015)\citenamefont {Huang}, \citenamefont {Zhang}, \citenamefont {Zou}, \citenamefont {Zou},\ and\ \citenamefont {Guo}}]{PhysRevA.91.043642}%
  \BibitemOpen
  \bibfield  {author} {\bibinfo {author} {\bibfnamefont {W.}~\bibnamefont {Huang}}, \bibinfo {author} {\bibfnamefont {Y.-L.}\ \bibnamefont {Zhang}}, \bibinfo {author} {\bibfnamefont {C.-L.}\ \bibnamefont {Zou}}, \bibinfo {author} {\bibfnamefont {X.-B.}\ \bibnamefont {Zou}},\ and\ \bibinfo {author} {\bibfnamefont {G.-C.}\ \bibnamefont {Guo}},\ }\bibfield  {title} {\bibinfo {title} {Two-axis spin squeezing of two-component {Bose-Einstein} condensates via continuous driving},\ }\href {https://doi.org/10.1103/PhysRevA.91.043642} {\bibfield  {journal} {\bibinfo  {journal} {Phys. Rev. A}\ }\textbf {\bibinfo {volume} {91}},\ \bibinfo {pages} {043642} (\bibinfo {year} {2015})}\BibitemShut {NoStop}%
\bibitem [{\citenamefont {Oka}\ and\ \citenamefont {Kitamura}(2019)}]{annurev}%
  \BibitemOpen
  \bibfield  {author} {\bibinfo {author} {\bibfnamefont {T.}~\bibnamefont {Oka}}\ and\ \bibinfo {author} {\bibfnamefont {S.}~\bibnamefont {Kitamura}},\ }\bibfield  {title} {\bibinfo {title} {Floquet engineering of quantum materials},\ }\href {https://doi.org/https://doi.org/10.1146/annurev-conmatphys-031218-013423} {\bibfield  {journal} {\bibinfo  {journal} {Annual Review of Condensed Matter Physics}\ }\textbf {\bibinfo {volume} {10}},\ \bibinfo {pages} {387} (\bibinfo {year} {2019})}\BibitemShut {NoStop}%
\bibitem [{\citenamefont {Swingle}\ \emph {et~al.}(2016)\citenamefont {Swingle}, \citenamefont {Bentsen}, \citenamefont {Schleier-Smith},\ and\ \citenamefont {Hayden}}]{PhysRevA.94.040302}%
  \BibitemOpen
  \bibfield  {author} {\bibinfo {author} {\bibfnamefont {B.}~\bibnamefont {Swingle}}, \bibinfo {author} {\bibfnamefont {G.}~\bibnamefont {Bentsen}}, \bibinfo {author} {\bibfnamefont {M.}~\bibnamefont {Schleier-Smith}},\ and\ \bibinfo {author} {\bibfnamefont {P.}~\bibnamefont {Hayden}},\ }\bibfield  {title} {\bibinfo {title} {Measuring the scrambling of quantum information},\ }\href {https://doi.org/10.1103/PhysRevA.94.040302} {\bibfield  {journal} {\bibinfo  {journal} {Phys. Rev. A}\ }\textbf {\bibinfo {volume} {94}},\ \bibinfo {pages} {040302} (\bibinfo {year} {2016})}\BibitemShut {NoStop}%
\bibitem [{\citenamefont {Li}\ \emph {et~al.}(2017)\citenamefont {Li}, \citenamefont {Fan}, \citenamefont {Wang}, \citenamefont {Ye}, \citenamefont {Zeng}, \citenamefont {Zhai}, \citenamefont {Peng},\ and\ \citenamefont {Du}}]{PhysRevX.7.031011}%
  \BibitemOpen
  \bibfield  {author} {\bibinfo {author} {\bibfnamefont {J.}~\bibnamefont {Li}}, \bibinfo {author} {\bibfnamefont {R.}~\bibnamefont {Fan}}, \bibinfo {author} {\bibfnamefont {H.}~\bibnamefont {Wang}}, \bibinfo {author} {\bibfnamefont {B.}~\bibnamefont {Ye}}, \bibinfo {author} {\bibfnamefont {B.}~\bibnamefont {Zeng}}, \bibinfo {author} {\bibfnamefont {H.}~\bibnamefont {Zhai}}, \bibinfo {author} {\bibfnamefont {X.}~\bibnamefont {Peng}},\ and\ \bibinfo {author} {\bibfnamefont {J.}~\bibnamefont {Du}},\ }\bibfield  {title} {\bibinfo {title} {Measuring out-of-time-order correlators on a nuclear magnetic resonance quantum simulator},\ }\href {https://doi.org/10.1103/PhysRevX.7.031011} {\bibfield  {journal} {\bibinfo  {journal} {Phys. Rev. X}\ }\textbf {\bibinfo {volume} {7}},\ \bibinfo {pages} {031011} (\bibinfo {year} {2017})}\BibitemShut {NoStop}%
\bibitem [{\citenamefont {G\"arttner}\ \emph {et~al.}(2018)\citenamefont {G\"arttner}, \citenamefont {Hauke},\ and\ \citenamefont {Rey}}]{PhysRevLett.120.040402}%
  \BibitemOpen
  \bibfield  {author} {\bibinfo {author} {\bibfnamefont {M.}~\bibnamefont {G\"arttner}}, \bibinfo {author} {\bibfnamefont {P.}~\bibnamefont {Hauke}},\ and\ \bibinfo {author} {\bibfnamefont {A.~M.}\ \bibnamefont {Rey}},\ }\bibfield  {title} {\bibinfo {title} {Relating out-of-time-order correlations to entanglement via multiple-quantum coherences},\ }\href {https://doi.org/10.1103/PhysRevLett.120.040402} {\bibfield  {journal} {\bibinfo  {journal} {Phys. Rev. Lett.}\ }\textbf {\bibinfo {volume} {120}},\ \bibinfo {pages} {040402} (\bibinfo {year} {2018})}\BibitemShut {NoStop}%
\bibitem [{\citenamefont {Lewis-Swan}\ \emph {et~al.}(2019)\citenamefont {Lewis-Swan}, \citenamefont {Safavi-Naini}, \citenamefont {Bollinger},\ and\ \citenamefont {Rey}}]{lewis2019unifying}%
  \BibitemOpen
  \bibfield  {author} {\bibinfo {author} {\bibfnamefont {R.~J.}\ \bibnamefont {Lewis-Swan}}, \bibinfo {author} {\bibfnamefont {A.}~\bibnamefont {Safavi-Naini}}, \bibinfo {author} {\bibfnamefont {J.~J.}\ \bibnamefont {Bollinger}},\ and\ \bibinfo {author} {\bibfnamefont {A.~M.}\ \bibnamefont {Rey}},\ }\bibfield  {title} {\bibinfo {title} {Unifying scrambling, thermalization and entanglement through measurement of fidelity out-of-time-order correlators in the {Dicke} model},\ }\href {https://doi.org/10.1038/s41467-019-09436-y} {\bibfield  {journal} {\bibinfo  {journal} {Nature communications}\ }\textbf {\bibinfo {volume} {10}},\ \bibinfo {pages} {1581} (\bibinfo {year} {2019})}\BibitemShut {NoStop}%
\bibitem [{\citenamefont {Landsman}\ \emph {et~al.}(2019)\citenamefont {Landsman}, \citenamefont {Figgatt}, \citenamefont {Schuster}, \citenamefont {Linke}, \citenamefont {Yoshida}, \citenamefont {Yao},\ and\ \citenamefont {Monroe}}]{landsman2019verified}%
  \BibitemOpen
  \bibfield  {author} {\bibinfo {author} {\bibfnamefont {K.~A.}\ \bibnamefont {Landsman}}, \bibinfo {author} {\bibfnamefont {C.}~\bibnamefont {Figgatt}}, \bibinfo {author} {\bibfnamefont {T.}~\bibnamefont {Schuster}}, \bibinfo {author} {\bibfnamefont {N.~M.}\ \bibnamefont {Linke}}, \bibinfo {author} {\bibfnamefont {B.}~\bibnamefont {Yoshida}}, \bibinfo {author} {\bibfnamefont {N.~Y.}\ \bibnamefont {Yao}},\ and\ \bibinfo {author} {\bibfnamefont {C.}~\bibnamefont {Monroe}},\ }\bibfield  {title} {\bibinfo {title} {Verified quantum information scrambling},\ }\href {https://doi.org/10.1038/s41586-019-0952-6} {\bibfield  {journal} {\bibinfo  {journal} {Nature}\ }\textbf {\bibinfo {volume} {567}},\ \bibinfo {pages} {61} (\bibinfo {year} {2019})}\BibitemShut {NoStop}%
\bibitem [{\citenamefont {Lewis-Swan}\ \emph {et~al.}(2020)\citenamefont {Lewis-Swan}, \citenamefont {Muleady},\ and\ \citenamefont {Rey}}]{PhysRevLett.125.240605}%
  \BibitemOpen
  \bibfield  {author} {\bibinfo {author} {\bibfnamefont {R.~J.}\ \bibnamefont {Lewis-Swan}}, \bibinfo {author} {\bibfnamefont {S.~R.}\ \bibnamefont {Muleady}},\ and\ \bibinfo {author} {\bibfnamefont {A.~M.}\ \bibnamefont {Rey}},\ }\bibfield  {title} {\bibinfo {title} {Detecting out-of-time-order correlations via quasiadiabatic echoes as a tool to reveal quantum coherence in equilibrium quantum phase transitions},\ }\href {https://doi.org/10.1103/PhysRevLett.125.240605} {\bibfield  {journal} {\bibinfo  {journal} {Phys. Rev. Lett.}\ }\textbf {\bibinfo {volume} {125}},\ \bibinfo {pages} {240605} (\bibinfo {year} {2020})}\BibitemShut {NoStop}%
\bibitem [{\citenamefont {Joshi}\ \emph {et~al.}(2020)\citenamefont {Joshi}, \citenamefont {Elben}, \citenamefont {Vermersch}, \citenamefont {Brydges}, \citenamefont {Maier}, \citenamefont {Zoller}, \citenamefont {Blatt},\ and\ \citenamefont {Roos}}]{PhysRevLett.124.240505}%
  \BibitemOpen
  \bibfield  {author} {\bibinfo {author} {\bibfnamefont {M.~K.}\ \bibnamefont {Joshi}}, \bibinfo {author} {\bibfnamefont {A.}~\bibnamefont {Elben}}, \bibinfo {author} {\bibfnamefont {B.}~\bibnamefont {Vermersch}}, \bibinfo {author} {\bibfnamefont {T.}~\bibnamefont {Brydges}}, \bibinfo {author} {\bibfnamefont {C.}~\bibnamefont {Maier}}, \bibinfo {author} {\bibfnamefont {P.}~\bibnamefont {Zoller}}, \bibinfo {author} {\bibfnamefont {R.}~\bibnamefont {Blatt}},\ and\ \bibinfo {author} {\bibfnamefont {C.~F.}\ \bibnamefont {Roos}},\ }\bibfield  {title} {\bibinfo {title} {Quantum information scrambling in a trapped-ion quantum simulator with tunable range interactions},\ }\href {https://doi.org/10.1103/PhysRevLett.124.240505} {\bibfield  {journal} {\bibinfo  {journal} {Phys. Rev. Lett.}\ }\textbf {\bibinfo {volume} {124}},\ \bibinfo {pages} {240505} (\bibinfo {year} {2020})}\BibitemShut {NoStop}%
\bibitem [{\citenamefont {Braum{\"u}ller}\ \emph {et~al.}(2022)\citenamefont {Braum{\"u}ller}, \citenamefont {Karamlou}, \citenamefont {Yanay}, \citenamefont {Kannan}, \citenamefont {Kim}, \citenamefont {Kjaergaard}, \citenamefont {Melville}, \citenamefont {Niedzielski}, \citenamefont {Sung}, \citenamefont {Veps{\"a}l{\"a}inen} \emph {et~al.}}]{braumuller2022probing}%
  \BibitemOpen
  \bibfield  {author} {\bibinfo {author} {\bibfnamefont {J.}~\bibnamefont {Braum{\"u}ller}}, \bibinfo {author} {\bibfnamefont {A.~H.}\ \bibnamefont {Karamlou}}, \bibinfo {author} {\bibfnamefont {Y.}~\bibnamefont {Yanay}}, \bibinfo {author} {\bibfnamefont {B.}~\bibnamefont {Kannan}}, \bibinfo {author} {\bibfnamefont {D.}~\bibnamefont {Kim}}, \bibinfo {author} {\bibfnamefont {M.}~\bibnamefont {Kjaergaard}}, \bibinfo {author} {\bibfnamefont {A.}~\bibnamefont {Melville}}, \bibinfo {author} {\bibfnamefont {B.~M.}\ \bibnamefont {Niedzielski}}, \bibinfo {author} {\bibfnamefont {Y.}~\bibnamefont {Sung}}, \bibinfo {author} {\bibfnamefont {A.}~\bibnamefont {Veps{\"a}l{\"a}inen}}, \emph {et~al.},\ }\bibfield  {title} {\bibinfo {title} {Probing quantum information propagation with out-of-time-ordered correlators},\ }\href {https://doi.org/10.1038/s41567-021-01430-w} {\bibfield  {journal} {\bibinfo  {journal} {Nature Physics}\ }\textbf {\bibinfo {volume} {18}},\ \bibinfo {pages} {172} (\bibinfo {year} {2022})}\BibitemShut
  {NoStop}%
\bibitem [{\citenamefont {Li}\ \emph {et~al.}(2025)\citenamefont {Li}, \citenamefont {Zhou}, \citenamefont {Zhang}, \citenamefont {Wu}, \citenamefont {Zhao}, \citenamefont {Yin}, \citenamefont {An}, \citenamefont {Zhai}, \citenamefont {Zhang}, \citenamefont {Peng},\ and\ \citenamefont {Du}}]{li2025error}%
  \BibitemOpen
  \bibfield  {author} {\bibinfo {author} {\bibfnamefont {Y.-C.}\ \bibnamefont {Li}}, \bibinfo {author} {\bibfnamefont {T.-G.}\ \bibnamefont {Zhou}}, \bibinfo {author} {\bibfnamefont {S.}~\bibnamefont {Zhang}}, \bibinfo {author} {\bibfnamefont {Z.}~\bibnamefont {Wu}}, \bibinfo {author} {\bibfnamefont {L.}~\bibnamefont {Zhao}}, \bibinfo {author} {\bibfnamefont {H.}~\bibnamefont {Yin}}, \bibinfo {author} {\bibfnamefont {X.}~\bibnamefont {An}}, \bibinfo {author} {\bibfnamefont {H.}~\bibnamefont {Zhai}}, \bibinfo {author} {\bibfnamefont {P.}~\bibnamefont {Zhang}}, \bibinfo {author} {\bibfnamefont {X.}~\bibnamefont {Peng}},\ and\ \bibinfo {author} {\bibfnamefont {J.}~\bibnamefont {Du}},\ }\href {https://arxiv.org/abs/2506.19915} {\bibinfo {title} {Error-resilient reversal of quantum chaotic dynamics enabled by scramblons}} (\bibinfo {year} {2025}),\ \Eprint {https://arxiv.org/abs/2506.19915} {arXiv:2506.19915} \BibitemShut {NoStop}%
\bibitem [{\citenamefont {Leroux}\ \emph {et~al.}(2010)\citenamefont {Leroux}, \citenamefont {Schleier-Smith},\ and\ \citenamefont {Vuleti\ifmmode~\acute{c}\else \'{c}\fi{}}}]{PhysRevLett.104.073602}%
  \BibitemOpen
  \bibfield  {author} {\bibinfo {author} {\bibfnamefont {I.~D.}\ \bibnamefont {Leroux}}, \bibinfo {author} {\bibfnamefont {M.~H.}\ \bibnamefont {Schleier-Smith}},\ and\ \bibinfo {author} {\bibfnamefont {V.}~\bibnamefont {Vuleti\ifmmode~\acute{c}\else \'{c}\fi{}}},\ }\bibfield  {title} {\bibinfo {title} {Implementation of cavity squeezing of a collective atomic spin},\ }\href {https://doi.org/10.1103/PhysRevLett.104.073602} {\bibfield  {journal} {\bibinfo  {journal} {Phys. Rev. Lett.}\ }\textbf {\bibinfo {volume} {104}},\ \bibinfo {pages} {073602} (\bibinfo {year} {2010})}\BibitemShut {NoStop}%
\bibitem [{\citenamefont {Greve}\ \emph {et~al.}(2022)\citenamefont {Greve}, \citenamefont {Luo}, \citenamefont {Wu},\ and\ \citenamefont {Thompson}}]{greve2022entanglement}%
  \BibitemOpen
  \bibfield  {author} {\bibinfo {author} {\bibfnamefont {G.~P.}\ \bibnamefont {Greve}}, \bibinfo {author} {\bibfnamefont {C.}~\bibnamefont {Luo}}, \bibinfo {author} {\bibfnamefont {B.}~\bibnamefont {Wu}},\ and\ \bibinfo {author} {\bibfnamefont {J.~K.}\ \bibnamefont {Thompson}},\ }\bibfield  {title} {\bibinfo {title} {Entanglement-enhanced matter-wave interferometry in a high-finesse cavity},\ }\href {https://doi.org/10.1038/s41586-022-05197-9} {\bibfield  {journal} {\bibinfo  {journal} {Nature}\ }\textbf {\bibinfo {volume} {610}},\ \bibinfo {pages} {472} (\bibinfo {year} {2022})}\BibitemShut {NoStop}%
\bibitem [{\citenamefont {Bohnet}\ \emph {et~al.}(2016)\citenamefont {Bohnet}, \citenamefont {Sawyer}, \citenamefont {Britton}, \citenamefont {Wall}, \citenamefont {Rey}, \citenamefont {Foss-Feig},\ and\ \citenamefont {Bollinger}}]{bohnet2016quantum}%
  \BibitemOpen
  \bibfield  {author} {\bibinfo {author} {\bibfnamefont {J.~G.}\ \bibnamefont {Bohnet}}, \bibinfo {author} {\bibfnamefont {B.~C.}\ \bibnamefont {Sawyer}}, \bibinfo {author} {\bibfnamefont {J.~W.}\ \bibnamefont {Britton}}, \bibinfo {author} {\bibfnamefont {M.~L.}\ \bibnamefont {Wall}}, \bibinfo {author} {\bibfnamefont {A.~M.}\ \bibnamefont {Rey}}, \bibinfo {author} {\bibfnamefont {M.}~\bibnamefont {Foss-Feig}},\ and\ \bibinfo {author} {\bibfnamefont {J.~J.}\ \bibnamefont {Bollinger}},\ }\bibfield  {title} {\bibinfo {title} {Quantum spin dynamics and entanglement generation with hundreds of trapped ions},\ }\href {https://doi.org/10.1126/science.aad9958} {\bibfield  {journal} {\bibinfo  {journal} {Science}\ }\textbf {\bibinfo {volume} {352}},\ \bibinfo {pages} {1297} (\bibinfo {year} {2016})}\BibitemShut {NoStop}%
\bibitem [{\citenamefont {Hines}\ \emph {et~al.}(2023)\citenamefont {Hines}, \citenamefont {Rajagopal}, \citenamefont {Moreau}, \citenamefont {Wahrman}, \citenamefont {Lewis}, \citenamefont {Markovi\ifmmode~\acute{c}\else \'{c}\fi{}},\ and\ \citenamefont {Schleier-Smith}}]{PhysRevLett.131.063401}%
  \BibitemOpen
  \bibfield  {author} {\bibinfo {author} {\bibfnamefont {J.~A.}\ \bibnamefont {Hines}}, \bibinfo {author} {\bibfnamefont {S.~V.}\ \bibnamefont {Rajagopal}}, \bibinfo {author} {\bibfnamefont {G.~L.}\ \bibnamefont {Moreau}}, \bibinfo {author} {\bibfnamefont {M.~D.}\ \bibnamefont {Wahrman}}, \bibinfo {author} {\bibfnamefont {N.~A.}\ \bibnamefont {Lewis}}, \bibinfo {author} {\bibfnamefont {O.}~\bibnamefont {Markovi\ifmmode~\acute{c}\else \'{c}\fi{}}},\ and\ \bibinfo {author} {\bibfnamefont {M.}~\bibnamefont {Schleier-Smith}},\ }\bibfield  {title} {\bibinfo {title} {Spin squeezing by {Rydberg} dressing in an array of atomic ensembles},\ }\href {https://doi.org/10.1103/PhysRevLett.131.063401} {\bibfield  {journal} {\bibinfo  {journal} {Phys. Rev. Lett.}\ }\textbf {\bibinfo {volume} {131}},\ \bibinfo {pages} {063401} (\bibinfo {year} {2023})}\BibitemShut {NoStop}%
\bibitem [{\citenamefont {Zhang}\ \emph {et~al.}(2025)\citenamefont {Zhang}, \citenamefont {Jin}, \citenamefont {Duan}, \citenamefont {M\o{}lmer}, \citenamefont {Zhang}, \citenamefont {Wang},\ and\ \citenamefont {Xiao}}]{xkt1-y58b}%
  \BibitemOpen
  \bibfield  {author} {\bibinfo {author} {\bibfnamefont {Y.}~\bibnamefont {Zhang}}, \bibinfo {author} {\bibfnamefont {S.}~\bibnamefont {Jin}}, \bibinfo {author} {\bibfnamefont {J.}~\bibnamefont {Duan}}, \bibinfo {author} {\bibfnamefont {K.}~\bibnamefont {M\o{}lmer}}, \bibinfo {author} {\bibfnamefont {G.}~\bibnamefont {Zhang}}, \bibinfo {author} {\bibfnamefont {M.}~\bibnamefont {Wang}},\ and\ \bibinfo {author} {\bibfnamefont {Y.}~\bibnamefont {Xiao}},\ }\bibfield  {title} {\bibinfo {title} {Cooperative squeezing of internal and collective spins in an atomic ensemble},\ }\href {https://doi.org/10.1103/xkt1-y58b} {\bibfield  {journal} {\bibinfo  {journal} {Phys. Rev. Lett.}\ }\textbf {\bibinfo {volume} {135}},\ \bibinfo {pages} {213604} (\bibinfo {year} {2025})}\BibitemShut {NoStop}%
\bibitem [{\citenamefont {Auccaise}\ \emph {et~al.}(2015)\citenamefont {Auccaise}, \citenamefont {Araujo-Ferreira}, \citenamefont {Sarthour}, \citenamefont {Oliveira}, \citenamefont {Bonagamba},\ and\ \citenamefont {Roditi}}]{PhysRevLett.114.043604}%
  \BibitemOpen
  \bibfield  {author} {\bibinfo {author} {\bibfnamefont {R.}~\bibnamefont {Auccaise}}, \bibinfo {author} {\bibfnamefont {A.~G.}\ \bibnamefont {Araujo-Ferreira}}, \bibinfo {author} {\bibfnamefont {R.~S.}\ \bibnamefont {Sarthour}}, \bibinfo {author} {\bibfnamefont {I.~S.}\ \bibnamefont {Oliveira}}, \bibinfo {author} {\bibfnamefont {T.~J.}\ \bibnamefont {Bonagamba}},\ and\ \bibinfo {author} {\bibfnamefont {I.}~\bibnamefont {Roditi}},\ }\bibfield  {title} {\bibinfo {title} {Spin squeezing in a quadrupolar nuclei {NMR} system},\ }\href {https://doi.org/10.1103/PhysRevLett.114.043604} {\bibfield  {journal} {\bibinfo  {journal} {Phys. Rev. Lett.}\ }\textbf {\bibinfo {volume} {114}},\ \bibinfo {pages} {043604} (\bibinfo {year} {2015})}\BibitemShut {NoStop}%
\end{thebibliography}
\end{document}